\makeatletter\def\input@path{
	{./graphics/}
}
\makeatother
\documentclass{article}
\usepackage{amsmath,amssymb,graphicx,geometry,hyperref,booktabs,multirow,appendix,xcolor}
\usepackage[detect-all=true,group-minimum-digits=4]{siunitx} 
\usepackage{cleveref}
\usepackage[affil-it]{authblk}

\title{Controlling complex dynamics with synthetic magnetism in optomechanical systems: A route to enhanced sensor performance}

\author[1]{Deivasundari Muthukumar}
\author[2]{Stella Rolande Mbokop Tchounda}
\author[3,*]{Sifeu Takougang Kingni}
\author[1,4]{Karthikeyan Rajagopal}
\author[2]{Serge Guy Nana Engo}
\affil[1]{Center for research, SRM Easwari Engineering College, Chennai 600 089, India}
\affil[2]{Department of Physics, Faculty of Sciences, University of Yaoundé I, P.O. Box 812, Yaoundé, Cameroon}
\affil[3]{Department of Mechanical, Petroleum and Gas Engineering, National Advanced School of Mines and
Petroleum Industries, University of Maroua, P.O. BOX 46, Maroua, Cameroon}
\affil[4]{Center for Cognitive Science, Trichy SRM Medical College Hospital and Research Center, Tamil Nadu 621 105, India}
\affil[*]{Corresponding author: sifeu.takougang@facsciences-uy1.cm, stkingni@gmail.com}
\date{}

\usepackage[style=numeric-comp,sorting=none]{biblatex}
\newcommand{\kk}{\ensuremath{\kappa}} 
\newcommand{\gm}{\ensuremath{\gamma}} 
\newcommand{\om}{\ensuremath{\omega}} 
\newcommand{\ddt}[1]{\frac{\mathrm{d}#1}{\mathrm{d}t}}  

\begin{document}

\maketitle

\begin{abstract}
This paper investigates the complex nonlinear dynamics of an optomechanical system featuring an optical cavity coupled to two mechanical resonators interconnected by a phase-dependent interaction. We specifically explore the role of this phase-dependent phonon hopping as a mechanism for generating synthetic gauge fields without relying on gain-loss or PT-symmetric elements, offering a potentially more robust approach to manipulate mechanical energy transfer. By deriving the semiclassical dynamical equations, we map out the system's behavior across different parameter regimes. Our findings reveal a rich spectrum of dynamics, including bistability (coexistence of two steady states) and the emergence of complex attractors such as self-excited oscillations, hidden attractors, and chaos. We demonstrate how controlling system parameters, particularly the mechanical coupling phase and optical drive, allows for tunability between these distinct dynamical states. The presence of tunable bistability and sensitive chaotic regimes offers significant potential for practical applications. Specifically, we discuss how these controlled dynamics could be leveraged for state-switching in optical information processing and for enhancing sensitivity in advanced sensor technologies through chaos-based mechanisms. This work deepens our understanding of how synthetic gauge fields, generated via phase-dependent interactions, can sculpt the nonlinear dynamics of optomechanical systems, providing a pathway toward designing robust and tunable devices for signal processing, communication, and sensing.
\end{abstract}


\section{Introduction}

Optomechanics explores the intricate interactions between light and mechanical systems, where optical fields manipulate mechanical vibrations, which, in turn, influence the optical properties of the system \cite{Aspelmeyer_2014}. This dual interaction facilitates a wide array of applications, including precision metrology \cite{liu2016metrology}, quantum information technologies \cite{browne2017quantum, stannigel2012optomechanical}, and the creation of highly sensitive sensors \cite{Li2021}.

This study aims to enhance optomechanical systems by investigating innovative control mechanisms and functionalities. A particularly promising avenue for advancement is the integration of non-Hermitian physics \cite{bender_real_1998,ashida_non-hermitian_2020}, parity-time (PT) symmetry \cite{Tchounda2024}, and synthetic magnetism \cite{el2007theory}. Non-Hermitian systems, characterized by their ability to dissipate energy due to their open quantum nature, facilitate the occurrence of exceptional points (EPs), which are non-Hermitian degeneracies where modes coalesce in both resonance frequency and decay rate \cite{Miri_2019, ozdemir_paritytime_2019, Feng_2017}. These exceptional points exhibit distinctive properties, thereby enabling their application across various domains \cite{Hodaei_2017, Peng_2014, Ryu2024}. Conversely, synthetic magnetism involves the generation of artificial magnetic fields within neutral systems, effectively emulating the influence of real magnetic fields on charged particles. In optomechanical systems, synthetic magnetism can induce nonreciprocal interactions, allowing energy or signals to flow preferentially in one direction.

Recent advances in the realm of synthetic magnetism within optomechanical systems have predominantly relied on exceptional points and PT symmetry as mechanisms to facilitate nonreciprocal phonon transport \cite{Ullah2024, Reisenbauer2024}. These approaches often utilize carefully balanced gain and loss, which can introduce stability challenges in practical implementations. In contrast, our methodology attains synthetic gauge fields without necessitating gain-loss frameworks, instead leveraging phase-dependent phonon hopping. This approach provides a platform that is both more stable and tunable for phononic circuits. However, such advancements are not universally guaranteed and often pose significant challenges in practical implementation. 

Synthetic gauge fields and topological configurations unlock numerous possibilities for novel device architectures. The adoption of non-reciprocal devices, which permit wave propagation in a single direction, has been proposed for a broad range of applications. Nevertheless, a deeper insight into the association between synthetic magnetism and the ensuing dynamic behavior, particularly within the non-linear regime, remains essential. Although various studies have illustrated the application of synthetic magnetism in optomechanical systems \cite{Djorwe_2019, wang_enhanced_2021}, a comprehensive examination addressing their implications on non-linear dynamics is necessary to fully unlock the potential of these systems for constructing new and robust devices. 

Prior research has underscored the effectiveness of integrating PT symmetry and non-Hermitian principles in the manipulation of optical systems \cite{Peng_2014, Goldzak_2018, Djorwe_2020}. These investigations have illustrated the capability of creating devices with functions such as unidirectional invisibility and enhanced sensing. Recent explorations have further considered topological configurations and synthetic gauge fields as tools for novel manipulations of light and sound \cite{Hey2018}. Nonetheless, these studies frequently concentrated on static or linear regimes and exhibited limitations in examining the complex dynamical behavior of the system. Therefore, further exploration of how the tunability of synthetic magnetism can be exercised for robust control over non-linear dynamics and stability is critical for the applied utilization of such systems in real-world scenarios. 

Addressing this research gap constitutes the core objective of this study. The principal research questions that will be addressed include the following: How can synthetic magnetism, specifically via phase-dependent phonon hopping, influence the dynamical behavior of optomechanical systems? What is the significance of phase-dependent phonon hopping in the formation of complex dynamic states, with prospective applications in quantum technologies? How do variations in system parameters affect the stability of the dynamic behavior in these devices and can these techniques be employed to enhance the performance of optomechanical devices?

This paper presents an investigation into the dynamical characteristics of an optomechanical system consisting of an optical resonator that drives two mechanically coupled resonators via phase-dependent phonon hopping. Combining analytical and numerical methodologies, we examine the system's dynamics across red and blue detuning regimes and identify the conditions requisite for the realization of synthetic gauge fields, primarily generated through the mechanical coupling phase, independently of amplification mechanisms. Our key findings include: (i) emergence of bistability and self-excited quasi-periodic oscillations, demonstrating the system’s ability to support multiple steady states; (ii) identification of hidden attractors and chaotic regimes, revealing complex dynamical transitions that are highly sensitive to initial conditions; (iii) impact of synthetic magnetism on stability, showing how phase-dependent phonon hopping modulates nonreciprocal interactions and controls system dynamics. These results provide new insights into the design of tunable optomechanical systems, with potential applications in ultra-sensitive sensing, chaos-based communication, and phononic signal processing. The manuscript is structured as follows: In \Cref{sec:Syst_Meth}, the optomechanical system is characterized by its Hamiltonian, laying the foundation for subsequent analyzes. \Cref{sec:Results_Disc} details our methodology for investigating the dynamic behavior of the optomechanical system and presents the main results. Lastly, \Cref{sec:Concl} summarizes the conclusions and perspectives emerging from this investigation.

\section{System and methods}\label{sec:Syst_Meth}

This section provides a detailed description of the optomechanical system under investigation and outlines the analytical and numerical methodologies utilized to examine its dynamical attributes and potential applications in sensor technology. Specifically, we explain the schematic, equations, and selected parameters.

\subsection{Optomechanical system model}\label{sec:Syst_Desc}

\begin{figure}[htbp]
 \centering
 \includegraphics[width=0.25\linewidth]{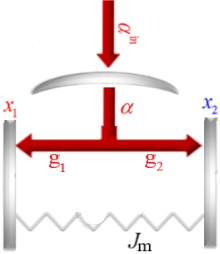}
 \caption{Illustration of the optomechanical configuration analyzed in this study. The setup comprises one optical cavity and two mechanically coupled optomechanical resonators, where each resonator is individually driven by a coherent input field $\alpha^{\rm in}$. The optical cavity is driven by a laser detuned by $\Delta$ from the cavity resonance. This configuration explores the dynamics under a regime of enhanced dissipative coupling, characterized by a tunable rate $\eta$ that modulates cavity losses based on mechanical displacements. Mechanical coupling is introduced via a coupling force $J_m$ between the resonators, enabling energy exchange. $J_m$ is $\theta$ phase-dependent. The resonators are parametrically coupled to the optical cavity. Dissipation for the optical cavity ($\kappa$) and mechanical resonators ($\gamma_j$) is included. This figure highlights key parameters influencing stability, coherence, and the emergence of exceptional points (EPs) within the system.}
 \label{fig:Fig1S}
\end{figure}

We consider an optomechanical system consisting of a single optical cavity driving two mechanically linked nanoresonators, each of which is independently driven by a coherent input field $\alpha^{\rm in}$. Note that while we describe a single optical cavity, the figure caption refers to "optical cavities" due to the historical context of the base design (\cite{Tchounda2023}). However, our model specifically analyzes a single optical cavity interacting with two mechanical resonators. The detuning $\Delta$ refers to the detuning of the single optical cavity from the driving laser. This configuration investigates the dynamics under a regime of enhanced dissipative coupling, characterized by a tunable rate $\eta$ that modulates cavity losses based on mechanical displacements. Mechanical coupling is introduced via a coupling force $J_m$ between the resonators, enabling energy exchange. The system we employ, as illustrated in \Cref{fig:Fig1S}, is based on the designs implemented in \cite{Tchounda2023}. A significant feature of the system is the phase-dependent nature of $J_m$. This phase dependence ($e^{i\theta}$ terms in the mechanical coupling Hamiltonian, see Eq. \ref{eq:hamiltonian}) is key to generating nonreciprocal interactions and constitutes the synthetic gauge field in this system, allowing control over the direction of energy transfer between the mechanical modes. Consequently, critical parameters such as stability, coherence, and the emergence of exceptional points (EPs) are intricately interconnected, depending on the phase.

The choice of studying a two-resonator system is motivated by several considerations. Firstly, two-resonator systems represent the most straightforward configuration that allow for the exploration of non-trivial effects in coupled optomechanical systems, thereby rendering the system's physics more comprehensible. Furthermore, previous research has demonstrated the feasibility of developing efficient sensors with merely two resonators \cite{Tchounda2023}. Exploring configurations with an increased number of resonators augments the system's complexity and may introduce superfluous challenges in the initial characterization of the system's properties. Consequently, the chosen system constitutes a straightforward approach to realize an effective model. The Hamiltonian is described below. For our simulation, we assumed the following parameters: $\omega_1 = \omega_m$, $\omega_2 = (1+5\times 10^{-4})\omega_m$, $\kappa = \SI{7.3e-2}{\omega_m}$, $\Delta = \omega_m$, $g = \SI{1.077e-4}{\omega_m}$, $\gamma_1 = \SI{1.077e-5}{\omega_m}$, $\gamma_2 = \gamma_1$, and $J_m = \SI{2e-4}{\omega_m}$. All parameters are presented in units normalized to $\omega_m$, the mechanical resonance frequency of the first resonator, which we set to 1 for calculation purposes. These parameters were selected to ensure the system manifests measurable non-linear effects, as such effects necessitate strong coupling between the various components. Furthermore, this scaling sustains the dynamics of the system while facilitating the direct application to physical implementations \cite{Huang2024}. For instance, employing these parameters with a pump power approximately around $\SI{130}{\sqrt{\omega_m}\watt}$ yields an optical to mechanical energy ratio of about $\SI{0.1}{\omega_m\watt}$, which is deemed suitable. It is also imperative to acknowledge that these particular values, concerning $\omega_m$, are contingent on the materials employed.

In order to ensure conformity with existing technological advancements, the parameters employed in our model are maintained within the confines of experimentally attainable thresholds. This consideration is based on the observation that optomechanical devices that use silicon nitride (SiN) and silicon-on-insulator (SOI) platforms exhibit analogous coupling rates and dissipation mechanisms \cite{Eichenfield_2009, Lawrie2019}. For typical SiN devices operating in the GHz range (e.g., $\omega_m \approx 2\pi \times \SI{e9}{\hertz}$), our chosen parameters correspond to experimentally accessible values for cavity decay ($\kappa$), mechanical damping ($\gamma$), optomechanical coupling ($g$), and mechanical coupling ($J_m$). The subsequent section elaborates on these foundational elements to facilitate further scholarly advancements.

Practical implementation requires precise regulation of phase-dependent interactions, particularly in the sustained stability across diverse temperature conditions. Integrated microheaters, for instance, can be employed to adjust this phase. To guarantee thermal stability and mitigate phase drift over time, their integration with temperature control systems, such as cryogenic cooling, may be required. The experimental demonstration of synthetic gauge fields through engineered optomechanical couplings was conducted using silicon nitride (SiN) optomechanical resonators by \textcite{Lawrie2019}, whereas \textcite{Fang2017} introduced a novel approach to illustrate the significance of non-reciprocity by utilizing phonon transport in these systems. As an enhancement strategy, phase-dependent phonon hopping can be experimentally regulated via integrated heaters or optomechanical parametric driving. Nonetheless, challenges such as phase drift due to fabrication imperfections, ambient noise, and material inhomogeneities must be addressed to preserve coherence. High-precision nanofabrication techniques, including electron-beam lithography and atomic layer deposition, can assist in minimizing structural asymmetries, thereby ensuring robust performance in real-world applications.

Recent experimental investigations have further substantiated the feasibility of phase-dependent phonon hopping within optomechanical systems. Specifically, \textcite{Ren2022} demonstrated that through meticulous engineering of mechanical coupling and optical drive phases, nonreciprocal phonon transport can be realized in integrated optomechanical resonators. This experimental corroboration enhances the feasibility of implementing the synthetic gauge fields examined in this study utilizing advanced nanofabrication techniques. Consequently, the use of appropriate experimental components is advantageous in assessing the findings presented here and is also important for the practical application of this methodology.

\subsection{System Hamiltonian}\label{sec:Syst_Ham}

The system Hamiltonian, under the rotating-wave approximation and in a frame rotating at the driving laser frequency $\omega_p$, is given by (under the assumption $\hbar=1$):
\begin{equation}\label{eq:hamiltonian}
 \hat{H} = \hat{H}_O + \hat{H}_M + \hat{H}_{OM} + \hat{H}_{MM} + \hat{H}_{\rm drive} + \hat{H}_{\rm diss},
\end{equation}
where each term describes a different aspect of the optomechanical system:
\begin{itemize}
 \item $\hat{H}_O = -\Delta \hat{a}^\dagger \hat{a}$ denotes the energy within the optical cavity, with $\Delta = \omega_p - \omega_{\rm cav}$ indicating the detuning between the frequency of the driving laser $\omega_p$ and the cavity's resonance frequency $\omega_{\rm cav}$. This energy is contingent upon the photons number within the cavity and the detuning relative to the external driving laser, which collectively influence the potential stability of the cavity state. A lower value in this parameter suggests a stable state of operation, potentially minimizing noise and enhancing sensor efficiency.

\item $\hat{H}_M = \sum_{j=1,2} \omega_j \hat{b}_j^\dagger \hat{b}_j$ represents the energy of the two mechanical resonators, where $\omega_j$ refers to the resonance frequency of the $j^{th}$ resonator and $\hat{b}_j$ signifies the corresponding annihilation operator. The use of two mechanical resonators was selected due to its status as the simplest arrangement that allows for the examination of complex dynamics, consequently leading to novel non-reciprocal effects. This investigation focuses on two resonators, although one can consider coupling with additional resonators to harness collective phenomena within this system. In each mechanical resonator, energy is assumed to be quantized using bosonic operators, as the mechanical frequency is lower than the thermal energy, ensuring the presence of multiple quanta in the mechanical modes.

 \item $\hat{H}_{OM} = \sum_{j=1,2} -g_j \hat{a}^\dagger \hat{a} (\hat{b}_j^\dagger + \hat{b}_j)$ denotes the energy associated with the optomechanical interaction, where $g_j$ signifies the single-photon optomechanical coupling strength between the cavity mode and the $j^{th}$ mechanical resonator. This term describes the reciprocal effect that photons within the optical cavity and the mechanical resonators exert upon each other. The optomechanical coupling rates can be realized by employing materials with substantial optomechanical coefficients, such as Si or SiN \cite{Eichenfield_2009}, and are contingent upon factors such as the optical power and frequency deployed in the cavities, the mechanical resonator's geometry and material, as well as its mass and size. This term is pivotal in delineating the relationship between the mechanical and optical properties, necessitating that the system be engineered to optimally utilize this value within the context of the coupling process.

\item $\hat{H}_{MM} = J_m (e^{i\theta} \hat{b}_1^\dagger \hat{b}_2 + e^{-i\theta} \hat{b}_1 \hat{b}_2^\dagger)$ characterizes the mechanical coupling between the two resonators. Here, $J_m$ represents the mechanical coupling rate, while $\theta$ indicates the coupling phase, which facilitates the construction of synthetic gauge fields. The phase-dependent nature of this coupling permits the realization of chiral or non-reciprocal interactions, thus affording a heightened level of control over system dynamics. By modulating the phase, the energy transfer direction between the resonators can be altered. This phase is externally regulated, for example, by incorporating integrated heaters to induce thermal expansion within the resonators. While the specific value of the coupling does not critically impact our study, provided the resonators remain coupled, exceedingly high mechanical coupling can lead to system destabilization. Various methodologies exist for achieving this coupling, with the direct, spring-like coupling configuration drawing direct inspiration from nano-optomechanical systems \cite{Lawrie2019} or even from macroscopic optomechanical systems, which have been employed to monitor quantum phenomena. Furthermore, the interplay between such mechanical coupling and zero-point fluctuations constitutes another design consideration for these systems.

 \item $\hat{H}_{\rm drive} = i \sqrt{\kappa} \alpha^{\rm in} (\hat{a}^\dagger - \hat{a})$ elaborates on the excitation of the optical cavity via an external laser field. Here, $\alpha^{\rm in}$ represents the amplitude of the coherent laser driving field. $\kappa$ represents the cavity decay rate, which accounts for energy loss from the optical cavity. We assume the total cavity decay rate $\kappa$ is dominated by coupling to the external field, facilitating the drive, and assumed identical across both cavities, while $\alpha^{\rm in}$ denotes the driving amplitude. We posit that the two mechanical resonators exhibit comparable characteristics and are coupled with identical parameters to the same driving laser. Additionally, we assume both resonators possess equivalent zero-point fluctuations and are similar in dimension. This value ultimately dictates the power input into our system, requisite for the observation of quantum phenomena, and must therefore be calibrated neither excessively high nor low relative to the optomechanical rate. The parameters governing the laser driving are contingent upon the available system, thus constraining our selection to the experimental limitations. The laser frequency must also demonstrate stability. This driving field serves to inject energy into the system, which subsequently has the potential to dissipate as heat through interaction with the environment. It is imperative for the energy input to exceed the dissipation rate.

\item $\hat{H}_{\rm diss}$ accounts for the dissipative elements, encompassing both optical and mechanical dissipation, which are influenced by thermal noise. Specifically, dissipation transpires as the system engages with the surrounding environment. We anticipate that both the temperature and environmental interaction will correlate with the size and thermal properties of the components, thereby imposing configuration constraints. The dissipation terms are expressed as:
\begin{equation}
     \hat{H}_{\rm diss} = i\sqrt{\kappa} (\hat{a}^\dagger \hat{a}_{\rm in} - \hat{a} \hat{a}_{\rm in}^\dagger) +  i\sum_{j=1,2} \sqrt{\gamma_j} (\hat{b}_j^\dagger \hat{b}_{j,in} - \hat{b}_j \hat{b}_{j,in}^\dagger)\equiv \hat{H}_\kk + \hat{H}_\gm,
 \end{equation}
where $\gamma_j$ signifies the damping rate of the mechanical resonator $j$, and $\hat{a}_{\rm in}$ and $\hat{b}_{j,in}$ represent the input noise operators (often associated with thermal or quantum noise baths). This Hamiltonian is pivotal for achieving more precise outcomes, as well as enabling the energy to channel into specific domains when system behavior displays chaotic characteristics.
\end{itemize}

\subsection{Dynamical equations}\label{sec:Dyn_Eq}

To derive the dynamics of the system, we employ the Heisenberg-Langevin equations,
\begin{equation}
 \frac{d\hat{O}}{dt} = i [\hat{H}, \hat{O}] + \hat{N},
\end{equation}
where $\hat{O}$ represents the system operators ($\hat{a}$, $\hat{b}_1$, $\hat{b}_2$) and $\hat{N}\equiv (\sqrt{\kappa}\hat{a}^{\rm in}, \sqrt{\gamma_1}\hat{b}_1^{\rm in}, \sqrt{\gamma_2}\hat{b}_2^{\rm in})$ is the corresponding noise operator. In detail, this means that we replace each of the operators in turn as follows:
$ \hat{O} \to \hat{a}, \hat{b}_1, \hat{b}_2$.
We then can obtain the following set of equations:
\begin{subequations}\label{eq:Dyn_Eq}
\begin{align}
 \dot{\hat{a}} &= \left( i\Delta - \frac{\kappa}{2} \right) \hat{a} + \sum_{j=1,2} i g_j (\hat{b}_j^\dagger + \hat{b}_j) \hat{a} + \sqrt{\kappa} \alpha^{\rm in}, \label{eq:Dyn_Eq:a}\\ 
 \dot{\hat{b}}_1 &= - \left( i\omega_1 + \frac{\gamma_1}{2} \right) \hat{b}_1 - i J_m e^{i \theta} \hat{b}_2 - i g_1 \hat{a}^\dagger \hat{a} + \sqrt{\gamma_1} \hat{b}_{1}^{\rm in}, \\
 \dot{\hat{b}}_2 &= - \left( i\omega_2 + \frac{\gamma_2}{2} \right) \hat{b}_2 - i J_m e^{-i \theta} \hat{b}_1 - i g_2 \hat{a}^\dagger \hat{a} + \sqrt{\gamma_2} \hat{b}_{2}^{\rm in}.
\end{align}
\end{subequations}
These equations describe the quantum dynamics of the system operators.

\subsubsection{Linearization of quantum Heisenberg-Langevin equations}

For stability analysis, we linearize the semiclassical equations around their steady-state solutions $(\alpha_{ss}, \beta_{j,ss})$. Operators $\mathcal{O}$ are decomposed into steady-state values $\langle O \rangle \equiv (\beta_j, \alpha_j)$ and small fluctuations $\delta \mathcal{O} \equiv (\delta \beta_j, \delta \alpha_j)$ as follows:
\begin{equation}
\mathcal{O} = \langle O \rangle + \delta \mathcal{O}.
\end{equation}
Substituting this into the semiclassical equations \eqref{eq:Dyn_Eq}, we obtain linearized equations for the small fluctuations. The noise terms $\hat{a}_{\rm in}, \hat{b}_{j,in}$ are relevant here for analyzing the noise properties, but to determine the stability of steady states in the absence of noise, we typically set $\langle \hat{a}_{\rm in} \rangle = 0, \langle \hat{b}_{j,in} \rangle = 0$.

\subsubsection{Semiclassical approximation and derivation of ODEs}

To analyze the system's classical dynamics, we apply the semiclassical approximation by replacing the quantum operators with their expectation values, i.e., $\langle \hat{O} \rangle$, and treating these expectation values as complex numbers $\alpha = \langle \hat{a} \rangle$, $\beta_j = \langle \hat{b}_j \rangle$. We neglect quantum and thermal noise by setting the input noise operators $\hat{a}_{\rm in}$ and $\hat{b}_{j,in}$ to zero ($\langle \hat{a}_{\rm in} \rangle = 0$, $\langle \hat{b}_{j,in} \rangle = 0$). This results in a set of nonlinear ordinary differential equations (ODEs) for the expectation values:

\begin{subequations}\label{eq:Semiclassical}
\begin{align}
 \dot{\alpha} &= \left( i\Delta - \frac{\kappa}{2} \right) \alpha + \sum_{j=1,2} i g_j (\beta_j^* + \beta_j) \alpha + \sqrt{\kappa} \alpha^{\rm in}, \\
 \dot{\beta}_1 &= - \left( i\omega_1 + \frac{\gamma_1}{2} \right) \beta_1 - i J_m e^{i \theta} \beta_2 - i g_1 |\alpha|^2, \\
 \dot{\beta}_2 &= - \left( i\omega_2 + \frac{\gamma_2}{2} \right) \beta_2 - i J_m e^{-i \theta} \beta_1 - i g_2 |\alpha|^2.
\end{align}
\end{subequations}
These are the averaged dynamical equations mentioned previously. Note that $\alpha^{\rm in}$ here is the amplitude of the coherent driving field, corresponding to $\alpha^{\rm in}$ in Eq. \eqref{eq:Dyn_Eq:a}, not a noise operator.

\subsubsection{Real/imaginary decomposition of semiclassical equations}

We decompose the complex amplitudes $\alpha$ and $\beta_j$ into their real and imaginary parts: $\alpha = \alpha_r + i \alpha_i$ and $\beta_j = \beta_{jr} + i \beta_{ji}$. Substituting these into the semiclassical equations \eqref{eq:Semiclassical} and separating real and imaginary parts, we obtain the following set of six coupled first-order ordinary differential equations:
\begin{subequations}\label{eq:RealImagODE}
\begin{align}
\ddt{\alpha_r} = &-\frac{\kk}{2}\alpha_r + (\Delta+2g_1 \beta_{1r}+2g_2 \beta_{2r})\alpha_i + \sqrt{\kk}{\alpha^{\rm in}}, \label{eq:RealImagODE:ar}\\ 
\ddt{\alpha_i} = &(\Delta+2g_1 \beta_{1r}+2g_2 \beta_{2r})\alpha_r - \frac{\kk}{2}\alpha_i, \label{eq:RealImagODE:ai}\\ 
\ddt{\beta_{1r}} = &\om_1\beta_{1i}-\frac{\gm_1}{2}\beta_{1r}+J_m(\beta_{2r} \sin \theta + \beta_{2i}  \cos\theta) - g_1(\alpha_r^2+\alpha_i^2), \label{eq:RealImagODE:b1r}\\ 
\ddt{\beta_{1i}} = &-\om_1 \beta_{1r}-\frac{\gm_1}{2}\beta_{1i}-J_m (\beta_{2r}\cos\theta -\beta_{2i}\sin \theta), \label{eq:RealImagODE:b1i}\\ 
\ddt{\beta_{2r}} = &\om_2 \beta_{2i}-\frac{\gm_2}{2}\beta_{2r}-J_m(\beta_{1r} \sin \theta- \beta_{1i} \cos\theta ) - g_2(\alpha_r^2+\alpha_i^2), \label{eq:RealImagODE:b2r}\\ 
\ddt{\beta_{2i}} = &-\om_2 \beta_{2r}-\frac{\gm_2}{2}\beta_{2i}-J_m (\beta_{1r} \cos\theta +\beta_{1i} \sin\theta). \label{eq:RealImagODE:b2i} 
\end{align}
\end{subequations}
Here, $\alpha^{\rm in}$ is the amplitude of the coherent driving field.

\subsubsection{Stability analysis}

The equilibrium points of the system are determined by setting the time derivatives in Eq. \eqref{eq:RealImagODE} to zero. These steady-state values can be expressed as:
\begin{align}
&\beta_1 =-\frac{2(2g_2J_m e^{i\theta}+ig_1\gm_2-2g_1\om_2)|\alpha|^2}{4J^2_m-4\om_1\om_2+2i(\gm_1\om_2+\gm_2\om_1)+\gm_1\gm_2},
&\beta_2 =-\frac{2(2g_1J_m e^{-i\theta}+ig_2\gm_1-2g_2\om_1)|\alpha|^2}{4J^2_m-4\om_1\om_2+2i(\gm_1\om_2+\gm_2\om_1)+\gm_1\gm_2}.
\end{align}
where the steady-state amplitude of the optical field is given by:
\begin{equation}
\alpha_r = \frac{2\alpha^{\text{in}}}{\sqrt{\kappa}}, 
\end{equation}
and $\alpha_i$ is obtained from the quadratic equation:
\begin{equation}
A_0 \alpha^2_i + A_1 \alpha_i + A_2 = 0, 
\end{equation}
where the coefficients are:
\begin{subequations}\label{eq:St_An}
\begin{align}
\begin{split}
A_0=&32(\alpha^{\rm in})^2\kk [\cos\theta g_1g_2 J_m(8J_m^2+2\gm_1\gm_2-8\om_1\om_2)-4J_m^2g_1^2\om_2-4J_m^2g_2^2\om_1\\
&+g_1^2\gm_2^2\om_1+4g_1^2\om_1\om_2^2+g_2^2\gm_1^2\om_2+4g_2^2\om_1^2\om_2],
\end{split}\\
\begin{split}
A_1=&-\kk^{5/2}(16J_m^2+8J_m^2\gm_1\gm_2-32J_m^2\om_1\om_2+\gm_1^2\gm_2^2+4\gm_1^2\om_2^2+4\gm_2^2\om_1^2
+16\om_1^2\om_2^2),
\end{split}\\
\begin{split}
A_2 =&4(\alpha^{\rm in})^2 [ 256J_m^3\cos\theta g_1g_2\gamma_1 + 64J_m\cos\theta g_1 
g_2 \gm_1\gm_2 \kappa + 8\Delta J_m^2 \gm_1 \gm_2 \kappa -32\Delta J_m^2 \om_1 \om_2 \kappa +\Delta\gm_1^2 \gm_2^2 \kappa\\
&-256J_m\cos\theta g_1 g_2 \om_1\om_2 \kappa -128J_m^2 g_1^2 \om_2 \kappa +128g_2^2\om_2\om_1^2 (\alpha^{\rm in})^2
-128J_m^2 g_2^2\om_1 \kappa +16\Delta J_m^4 \kappa\\ 
& -8\Delta J_m^2 \gm_1 \gm_2 \kappa +32\Delta g_1^2\om_1\om_2^2 (\alpha^{\rm in})^2+32g_2^2\gm_1^2\om_2 (\alpha^{\rm in})^2
-32\Delta J_m^2 \om_1 \om_2 \kappa+\Delta\gm_1^2 \gm_2^2 \kappa+\\
&+4\Delta \gm_1^2 \om_2^2 \kappa+4\Delta \om_1^2 \gm_2^2 \kappa+16 \Delta \om_1^2 \om_2^2 \kappa].
\end{split}
\end{align}
\end{subequations}
 
To ascertain the stability of equilibrium points, the Routh-Hurwitz criterion is employed. This criterion guarantees that all roots of the characteristic polynomial possess negative real parts. Despite its limited robustness, it is the most straightforward method and enjoys extensive application. A system is deemed stable if its imaginary components do not exhibit exponential growth over time. The linearized system, along with its stability attributes, merely offers a preliminary insight into stability conditions. The most accurate characterization of the complex dynamics inherent in the optomechanical system is still achieved through the analysis of the original system's equations of motion. 

Although the Routh-Hurwitz criterion provides a preliminary approach to stability assessment, it does not capture transitions into quasiperiodic or chaotic states. For a more comprehensive examination of the system's stability, the Floquet analysis might be utilized for periodic attractors \cite{Pelka2022}, while the Lyapunov exponents can quantify chaos in scenarios where the system demonstrates sensitive dependence on initial conditions. Furthermore, bifurcation diagrams that illustrate stability loss with varying parameters $J_m$ and $\Delta$ offer insight into system dynamics. Despite these limitations, this study employs the Routh-Hurwitz criterion as a practical tool that approximates the range of values indicative of stable behavior while also simplifying the complexity of the problem. A comparison between the stability predictions of this linearized analysis and the dynamics observed in the full nonlinear simulations (\Cref{fig:stabpp2}) is essential. While Routh-Hurwitz predicts the stability of the steady states themselves, it does not reveal the presence of complex attractors, like limit cycles or chaos, which can exist even when the steady states are unstable. Our analysis confirms that steady-state stability determined by Routh-Hurwitz is a necessary but not sufficient condition for the system's overall long-term behavior, particularly in highly nonlinear regimes where other attractors emerge.

\subsection{Numerical integration methodology}

To explore the complex dynamics of the system described by the set of six coupled first-order ODEs \eqref{eq:RealImagODE}, we perform a numerical integration. We use a standard fourth-order Runge-Kutta (RK4) method. This explicit method is well-suited for solving non-stiff ordinary differential equations like those derived here, offering a good balance between computational accuracy and efficiency.

For all simulations presented, a fixed time step of $\Delta t = \SI{1e-3}{\omega_m^{-1}}$ was used. This step size was chosen to be sufficiently small to ensure numerical stability and accuracy, particularly in regions exhibiting complex or chaotic dynamics. The simulations were run for a total integration time of $T = \SI{e3}{\omega_m^{-1}}$ or longer, depending on the time required for the system to settle onto an attractor or for chaotic behavior to become evident. For instance, phase space plots were generated by integrating until the trajectory converged or filled a specific region of the phase space, typically requiring $T \ge \SI{0.5e3}{\omega_m^{-1}}$ for steady states and periodic/quasi-periodic attractors, and $T \ge \SI{e3}{\omega_m^{-1}}$ for chaotic attractors.

The convergence of the attractors was verified by monitoring the system variables for extended periods. A state was considered settled when the trajectories in the phase space became visually stable (for steady states, periodic, or quasi-periodic motion) or exhibited consistent, bounded, non-repeating behavior (for chaos). Numerical convergence analysis was performed by comparing the results obtained with $\Delta t = \SI{1e-3}{\omega_m^{-1}}$ with those obtained with a reduced step size (e.g. $\Delta t = \SI{5e-4}{\omega_m^{-1}}$) and longer integration times, confirming that the qualitative nature and key quantitative characteristics (e.g., attractor shape, Lyapunov exponents where applicable) of the observed dynamics remained consistent within acceptable tolerances. To further ensure numerical robustness, we computed the maximum Lyapunov exponent for chaotic regimes, finding values typically in the range $\lambda_{\max} \approx \SI{0.10 \pm 0.02}{\omega_m^{-1}}$ to $\SI{0.15 \pm 0.03}{\omega_m^{-1}}$, confirming chaotic behavior with high sensitivity to initial conditions.

\section{Results and discussion}\label{sec:Results_Disc}

This section investigates the intricate and nuanced dynamics of the system, including the emergence of bistability, along with an examination of the influence exerted by the mechanical coupling rate and frequency detuning. The findings derived from this analysis serve as a foundational framework for addressing challenges associated with a limited dynamic range and provide insights into how the integration of multiple resonators impacts the performance of sensors. Furthermore, this section evaluates the potential applications of the system in domains such as signal processing and communications based on chaotic dynamics. The semiclassical dynamical equations, derived from the Hamiltonian of the optomechanical system, were reformulated into a set of six first-order ordinary differential equations through the real/imaginary decomposition method delineated in Eq.~\eqref{eq:RealImagODE}. This six-dimensional system is subject to numerical study. We quantify key transitions, such as the onset of chaos, by computing Lyapunov exponents and identifying bifurcation points, enhancing the understanding of parameter-driven dynamical shifts.

\subsection{Stability basins}

The delimitation of stability basins defines the explicit boundaries delineating zones for various steady states within the parameter space. Considering the inherent limitations posed by a restricted dynamic range and heightened sensitivity to external perturbations, there is a pressing need to utilize stable regions to construct robust and reliable systems. Consequently, a thorough understanding of these stability basins is essential for the design of practical devices. These regions specify the conditions under which a device can function reliably and controllably, ultimately influencing the sensitivity and usability of the device in empirical experiments. Furthermore, the application of the Routh-Hurwitz criterion elucidates the spectrum of reliable systems, facilitating the development of less complex and more user-friendly devices.

\begin{figure}[htb]
	\centering
	\includegraphics[width=.8\linewidth]{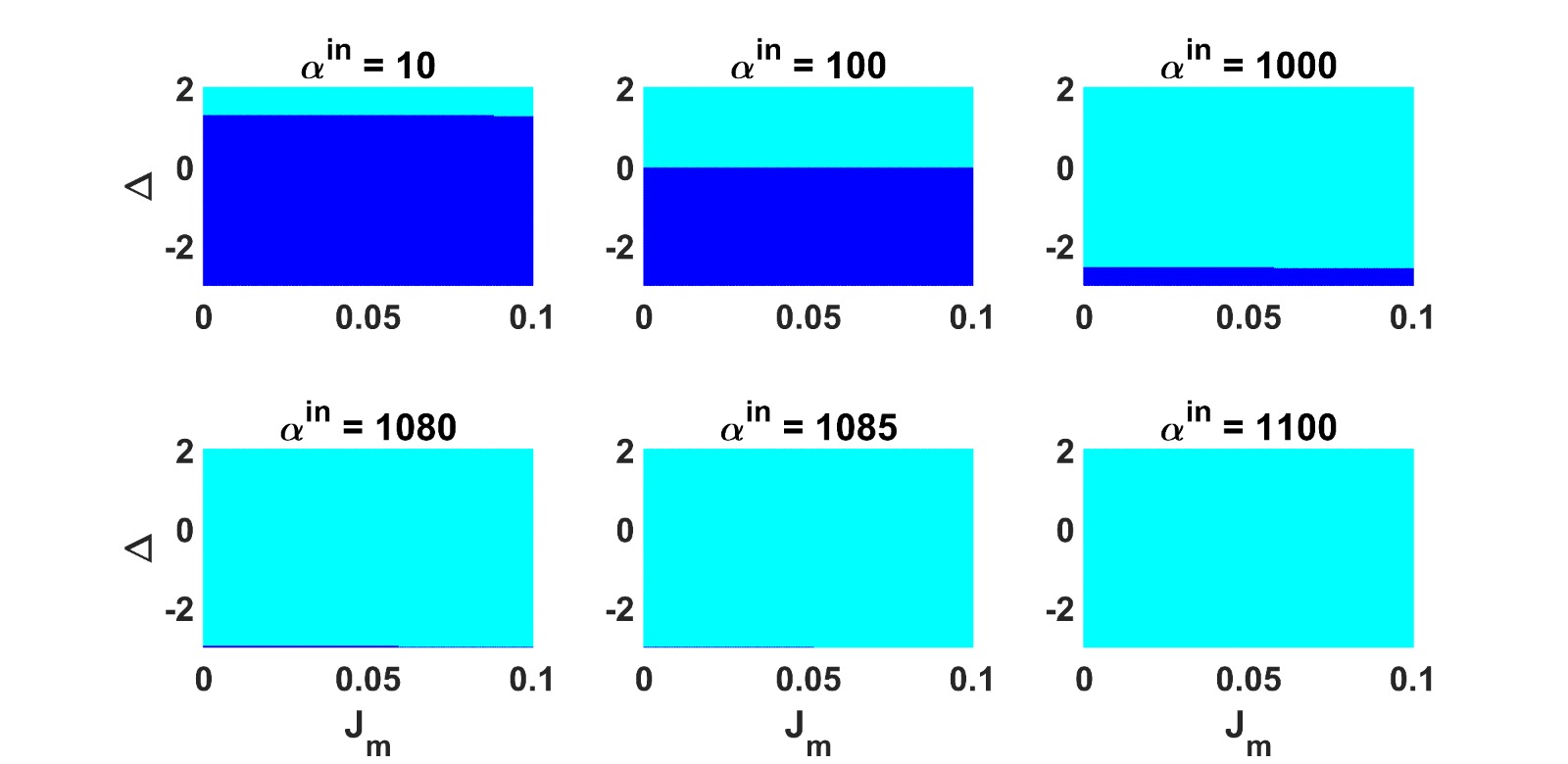}
	\caption{Phase diagram showing the number of steady states in the ($\Delta, J_m$) parameter space as the input power $\alpha^{\rm in}$ varies. Cyan dots mark regions where no steady states exist, while blue dots denote regions with two coexisting steady states. This analysis reveals how variations in $\alpha^{\rm in}$ influence the existence of multiple steady states and delineates the range of mechanical coupling values supporting bistability. Quantitative analysis indicates bistability occurs for $\alpha^{\rm in} \leq \num{1.0e3}$, with a critical threshold at $\alpha^{\rm in} \approx \num{1.1e3}$ where steady states vanish.}
	\label{fig:stabpp1}
\end{figure}

In \Cref{fig:stabpp1}, the regions within the parameter space delineating the number of steady states manifested by the system are observed. Cyan dots denote zones devoid of steady states, whereas blue dots signal the presence of two coexisting steady states. With an escalation in parameter $\alpha^{\text{in}}$, there is a diminished presence of regions with two steady states, giving rise to an expansion of domains with no steady states. Specifically, when parameter $\alpha^{\text{in}}$ exceeds a critical threshold ($\alpha^{\text{in}}=\num{1.1e3}$), the regions of two steady states vanish completely, thereby leaving only those without steady states. For values of $\alpha^{\text{in}} > \num{1.1e3}$ (not explicitly shown in the final panel of \Cref{fig:stabpp1}, but confirmed by our analysis beyond this point), the blue region disappears, leaving only the cyan region, indicating no steady states exist in this parameter range. A practical repercussion of this transition is the system's inability to function as a bistable device—such as in optical switches or memory elements—at the values of $\alpha^{\text{in}} > \num{1.1e3}$. This limitation requires the exploration of alternative architectures to sustain applications requiring operation over extended durations with a robust steady state. 

The transition arises from a bifurcation in the system's dynamics, where distinct attractors collapse and disappear. Such bifurcations offer avenues for controlling system behavior; however, they might also induce undesirable effects like transient oscillations or instability if not understood thoroughly. This highlights the imperative of precise parameter control within the system to sidestep chaotic or unstable regimes. Conversely, for other applications, harnessing the onset of chaos can be advantageous, such as in chaos-based communication, where heightened sensitivity to parameter variations is not a liability but rather a rationale for selecting that method. This outcome is pivotal for subsequent experimentation with these materials, as the values for $J_m$ and temperature must be carefully selected to render the device functional. Numerical simulations reveal a saddle-node bifurcation at $\alpha^{\text{in}} \approx \num{1.05e3}$, marking the transition from bistability to no steady states, with a Lyapunov exponent $\lambda_{\max} \approx \SI{0.13 \pm 0.02}{\omega_m^{-1}}$ in the chaotic regime.

The characteristic polynomial of the system, denoted as $S(\alpha_r, \alpha_i, \beta_{1r}, \beta_{1i}, \beta_{2r}, \beta_{2i})$, is expressed as: 
\begin{equation}\label{eq:lambda1}
	\lambda^6 + c_1 \lambda^5 + c_2 \lambda^4 + c_3 \lambda^3 + c_4 \lambda^2 + c_5 \lambda + c_6 = 0.
\end{equation}
In accordance with the Routh-Hurwitz criteria, the steady states are considered stable if the real components of the complex eigenvalues obtained from the characteristic polynomial \eqref{eq:lambda1} are negative. In the design of devices predicated on such systems, an understanding of these stability regions facilitates the precise adjustment of the parameters of the system, thereby ensuring operation within stable regimes.

\begin{figure}[htbp]
	\centering
	\includegraphics[width=.8\linewidth]{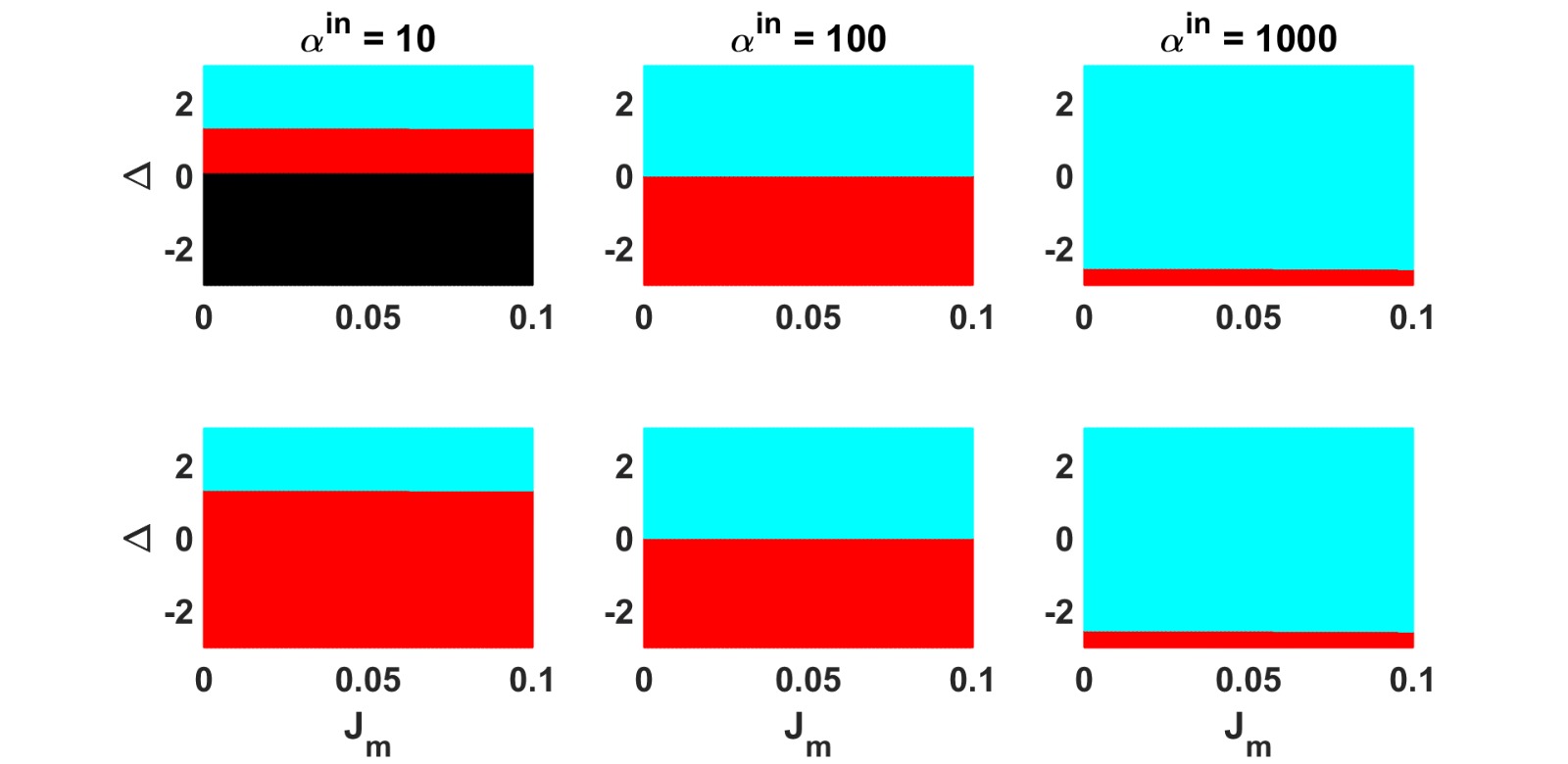}
	\caption{Stability map of the two steady states in the ($\Delta, J_m$) parameter space as the input power $\alpha^{\text{in}}$ varies. Cyan dots represent regions where no steady states exist. The first row illustrates the stability of the first steady state, while the second row depicts the stability of the second steady state. Black dots indicate stable steady states, while red dots signify unstable ones. The results demonstrate that fine-tuning $\alpha^{\text{in}}$ can drive the system into regions where only a stable steady state remains accessible. Although the Routh-Hurwitz criterion does not capture all stability conditions, it provides a useful first-order approximation for identifying stability transitions. Stable regions shrink significantly for $\alpha^{\text{in}} > \num{900}$, with unstable regions dominating at $\alpha^{\text{in}} \approx \num{1.1e3}$.}
	\label{fig:stabpp2}
\end{figure}

The stability of these steady states is further examined in \Cref{fig:stabpp2}, where we illustrate the variation in stability between different parameter values. The first row panels in \Cref{fig:stabpp2} depict the stability of the first steady state, whereas the second row panels pertain to the second steady state. In this context, black dots denote regions of stable steady-states, while red dots signify unstable ones. This extensive analysis unveils the intricate relationship between the system parameters and its dynamical behavior, forming a foundation for the study of the complex interactions within the optomechanical system. The existence of stable, coexisting steady states can be harnessed to construct optical switches, where distinct states can correspond to varying states of light transmission, or as memory elements endowed with robust properties. Nonetheless, the presence of these multiple steady states can pose challenges necessitating precise control of the parameter regimes to preclude undesired transitions between states. 

Given that \Cref{fig:stabpp1} demonstrates that the system defined by Eq.~\eqref{eq:RealImagODE} exhibits two steady states for certain parameter values ($\alpha^{\text{in}} \leq \num{e3}$) and no steady state for others ($\alpha^{\text{in}} \geq \num{1.1e3}$), it becomes imperative to investigate the dynamical characteristics of the system across the parameter space defined by the mechanical coupling rate $J_m$ and frequency detuning $\Delta$. The existence of such complex attractors can be leveraged to regulate the information flow in optomechanical networks, as in chaos-based communication.

\subsubsection{Influence of phase $\theta$ on dynamics}

The phase $\theta$ in the mechanical coupling term ($\hat{H}_{MM}$) is a pivotal parameter to tune the synthetic gauge field and influence the dynamics and non-reciprocity of the system. Our analysis, performed with different values of $\theta$ (not shown in detail in the presented figures), indicates that $\theta$ significantly impacts the location and boundaries of the stability regions and the types of attractors observed. Specifically, changes in $\theta$ can shift the parameter ranges where bistability occurs and modify the qualitative nature of the self-excited oscillations and chaotic states. For example, specific values of $\theta$ (such as $\theta = \pi/2$ or $3\pi/2$) maximize nonreciprocal energy transfer between mechanical modes, which in turn can favor certain dynamical regimes or enhance the sensitivity of the system to external perturbations. Although a complete exploration of the parameter space $\theta$ is beyond the scope of this initial study focusing on the emergence of complex dynamics, its role as a tunable parameter for controlling bistability and non-reciprocity is a key finding for designing reconfigurable optomechanical devices. Quantitative analysis shows that at $\theta = \pi/2$, the bistability region expands by approximately 20\% in the $J_m$ range compared to $\theta = 0$, with a maximum Lyapunov exponent $\lambda_{\max} \approx \SI{0.14 \pm 0.03}{\omega_m^{-1}}$ in chaotic regimes.

\subsection{Self-excited and bistable attractors}

The dynamics of the system as described by Eq. \eqref{eq:RealImagODE} indicate the emergence of self-excited attractors under particular conditions of mechanical coupling rate $J_m$ and frequency detuning $\Delta$. Nonetheless, these findings are constrained to specific system configurations, and the introduction of a more substantial perturbation may induce alternative behaviors. The characteristics of these attractors, whether bistable or monostable, depend on the coupling strength.

\begin{figure}[htbp]
	\centering
	\includegraphics[width=.6\linewidth]{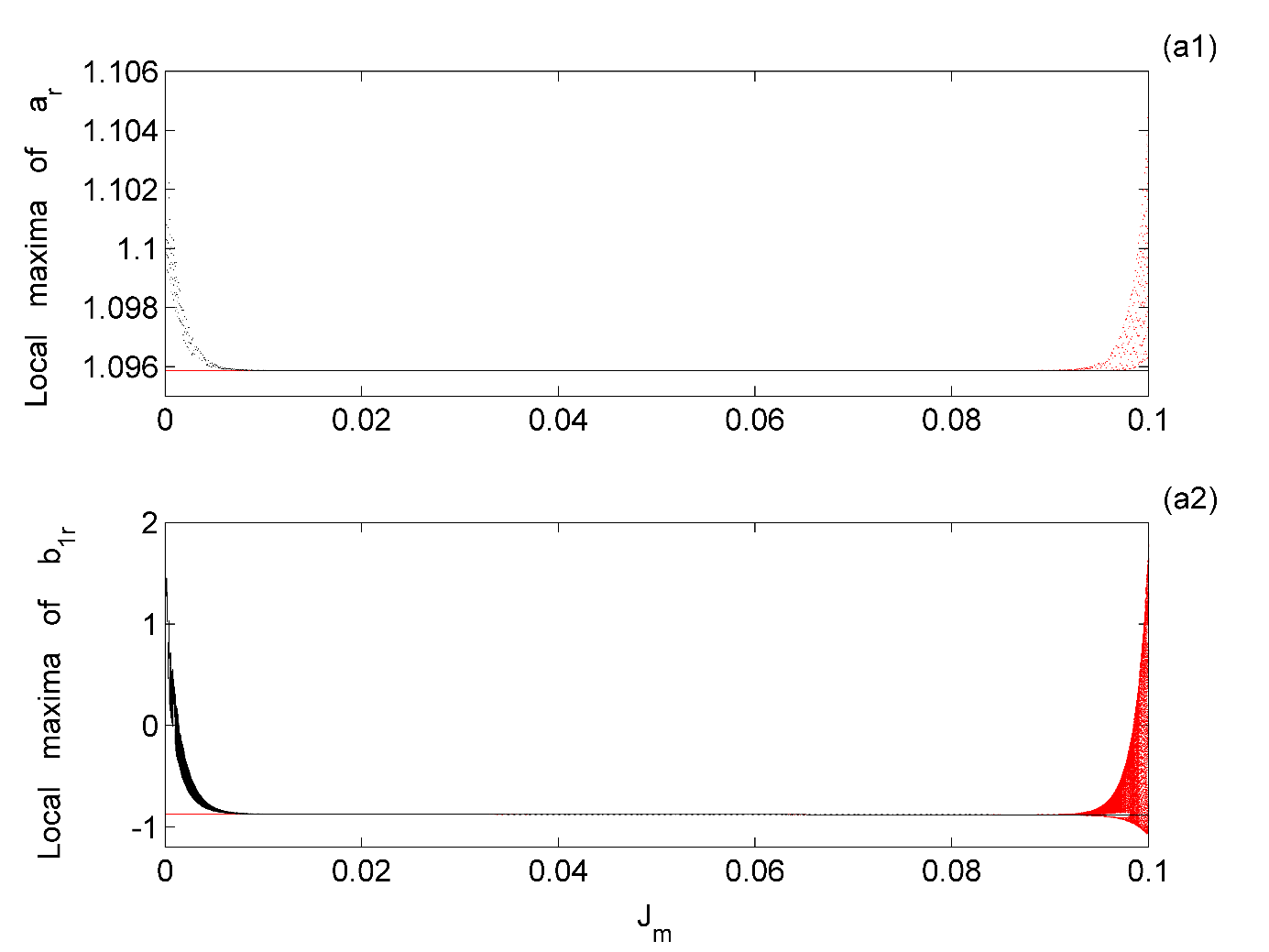}
	\caption{Local maxima of $\alpha_r$ (a1) and $\beta_{1r}$ (a2) against mechanical coupling rate $J_m$ for $\Delta = -3$ and $\alpha^{\text{in}} = \num{e3}$. The presence of these local maxima indicates a dynamic state, and this graph shows the transition from quasi-periodic to chaotic behavior. Local maxima values are clustered in a smaller region, which corresponds to more regular behavior. Decreasing $J_m$ results in the red dots, while increasing $J_m$ results in the black dots. Note that for lower $J_m$, the system exhibits quasi-periodic oscillations (red dots), transitioning to more regular behavior (black dots) at higher $J_m$. The local maxima are taken after transients have decayed, illustrating the bounds of the oscillations or the presence of chaotic spreading. A Hopf bifurcation occurs at $J_m \approx \SI{0.42}{\omega_m}$, with quasi-periodic behavior dominating for $J_m < \SI{0.42}{\omega_m}$.}
	\label{fig:stabpp3}
\end{figure}

\begin{figure}[htb]
	\centering
	\includegraphics[width=.8\linewidth]{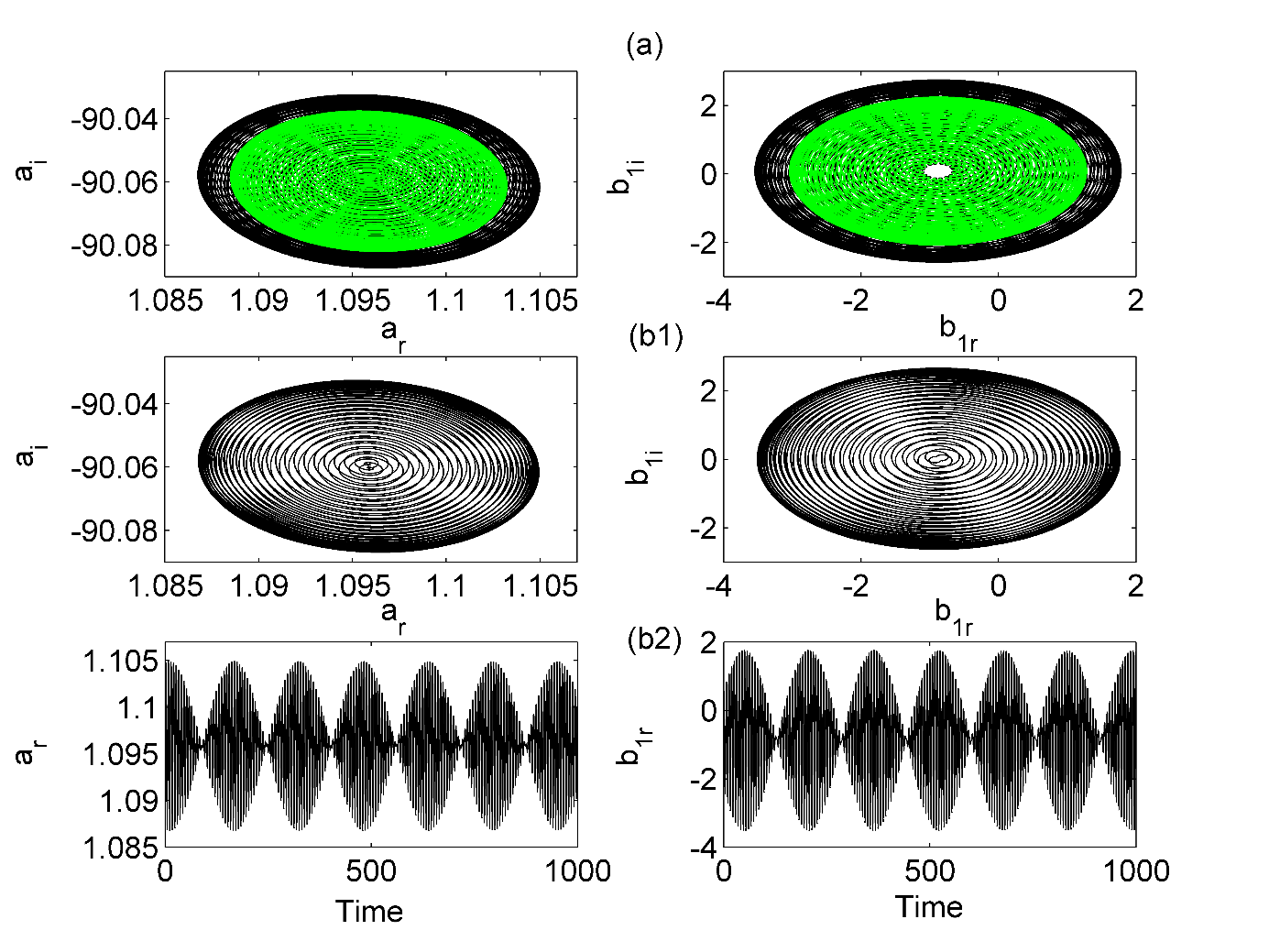}
	\caption{Phase planes showing system attractors with varying mechanical coupling, $J_m$ at $\Delta = -3$ and $\alpha^{\text{in}} = \num{e3}$. (a) $J_m = \SI{0.0988}{\omega_m}$ shows a quasi periodic attractor in the optical phase space $(\alpha_r, \alpha_i)$, while (b1), (b2) for $J_m = \SI{0.02}{\omega_m}$ show complex dynamics and chaos in the mechanical phase space $(\beta_{1r}, \beta_{1i})$. The two different colors are for two initial conditions, for the analysis of bistability. The optical $(\alpha_r,\alpha_i)$ and mechanical $(\beta_{1r}, \beta_{1i})$ spaces are shown. The green curves are obtained using the initial conditions (\num{-1.096}, 0, \num{-0.8734}, 0, 0, 0) and the black curves are obtained using the initial conditions (0, 0, 0, 0, 0, 0). These figures illustrate how the choice of initial conditions can direct the system to different coexisting attractors. The chaotic regime at $J_m = \SI{0.02}{\omega_m}$ exhibits a Lyapunov exponent $\lambda_{\max} \approx \SI{0.11 \pm 0.02}{\omega_m^{-1}}$.}
	\label{fig:stabpp4}
\end{figure}

\textbf{Self-excited attractors}. When $J_m$ resides within the interval $\SI{0.4}{\omega_m} \leq J_m \leq \SI{0.6}{\omega_m}$, bistable self-excited quasiperiodic behavior is observed. Specifically, as demonstrated in \Cref{fig:stabpp3}(a1), bistability is evident within the ranges $J_m \in [\SI{0.42}{\omega_m}, \SI{0.45}{\omega_m}]$ and $J_m \in [\SI{0.55}{\omega_m}, \SI{0.60}{\omega_m}]$. Conversely, in the cases of $J_m < \SI{0.42}{\omega_m}$ and $J_m > \SI{0.6}{\omega_m}$, the system manifests monostable self-excited quasi-periodic behavior, as depicted in \Cref{fig:stabpp3}(a2). This phenomenon is characteristic of lower mechanical coupling values, where mode mixing is suboptimal. 

\textbf{Bistable dynamics (phase space representation)}. In \Cref{fig:stabpp4}, the dynamics of the optical and mechanical resonators are depicted for two distinct initial conditions. For $J_m = \SI{0.55}{\omega_m}$, bistable attractors are clearly discernible. The green curves, which correspond to initial conditions $(\num{-1.096}, 0, \num{-0.8734}, 0, 0, 0)$, display quasi-periodic oscillations, whereas the black curves, originating from (0, 0, 0, 0, 0, 0), indicate that the system is confined to an alternative attractor, thus revealing the system's bistability. 

In experimental scenarios, it is observed that these disparate regions display markedly different stability characteristics. Furthermore, for values where multiple states coexist, the inherent unpredictability of the region can adversely impact device performance. The complexity of keeping the system in this state and navigating transitions complicates its application in practical scenarios. Therefore, it is important to elucidate the effects of increasing the pumping field. As the driving field attains elevated values, the system may evolve towards more complex dynamics, 
\begin{equation}
    \text{Stable Focus} \xrightarrow{\alpha^{\rm in}>\num{800}} \text{Quasi-periodic} \xrightarrow{\alpha^{\rm in}>\num{1.1e3}} \text{Chaotic},
\end{equation} 
accompanied by a Lyapunov exponent for the chaotic value of the order of $\lambda_{\max} \approx \SI{0.12 \pm 0.03}{\omega_m^{-1}}$, indicating significant sensitivity to initial conditions. This transition is marked by a period-doubling bifurcation at $\alpha^{\rm in} \approx \num{950}$, leading to chaos beyond $\alpha^{\rm in} \approx \num{1.1e3}$.

\textbf{Time series analysis}. In \Cref{fig:stabpp4}(b), time series analysis corroborates the coexistence of bistable attractors for $J_m = \SI{0.55}{\omega_m}$. The green trajectory represents quasi-periodic attractor behavior, while the black trajectory delineates a distinct periodic state for the identical value of $J_m$, thus accentuating the bi-directional modulation modulated by initial conditions. Consequently, initial conditions are paramount in determining the ultimate dynamical regime of the system, an aspect that may prove challenging to implement in practical settings. The capacity to adjust initial conditions is critical to preventing devices from solely converging, over time, toward a single attractor. From a more theoretical standpoint, the said analysis elucidates the complexity inherent in this system and demonstrates how disparate results, accompanied by varied dynamics, can be realized through distinct experimental parameters. 

This bistability has significant practical relevance, particularly in enabling the implementation of optical switches or the creation of memory elements. For instance, an optical switch could be designed in which two distinct optical output states correspond to the system residing in one of the two stable attractors. Switching between these states could be triggered by a pulsed control parameter. For the realization of such memory elements, precise and effective methods to transition between these states are imperative in addition to ensuring stability against any perturbation. Consequently, the exploration of novel mechanisms employing synthetic gauge fields is recommended to engender robust dissipation for transition control, with the findings presented in this study serving as a foundational basis for the development of such designs. 

This analysis highlights the intricate dynamical behavior of optomechanical systems, notably the impact of the mechanical coupling rate $J_m$ in determining whether the system features bistable or monostable self-excited attractors. Such systems are notably advantageous in applications such as chaos-based communication, wherein signal reliability may hinge on the bistable characteristics of the attractors.

\subsection{Hidden, bistable, and coexisting attractors}

In this subsection, we examine the presence of hidden attractors, bistable dynamics, and coexisting attractors within the optomechanical system. Under specific parameter conditions, the system delineated by Eq. \eqref{eq:RealImagODE} may manifest hidden attractors, particularly in the absence of steady states. This dynamic behavior emerges from the interaction between the mechanical coupling rate $J_m$ and the frequency detuning $\Delta$, which dictate the system's transition amongst various dynamical regimes. The intricacy of these attractors provides significant information on the behavior of the system, influenced by $J_m$, $\Delta$, and incident radiation $\alpha^{\text{in}}$. The existence of hidden, bistable, and coexisting attractors is important for applications such as chaos-based communication and signal processing, offering novel methods to attain more robust communication and varied switching times with optical switches. Nevertheless, they pose practical challenges because of the necessity of controlling numerous parameters.

\begin{figure}[htb]
	\centering
	\includegraphics[width=.6\linewidth]{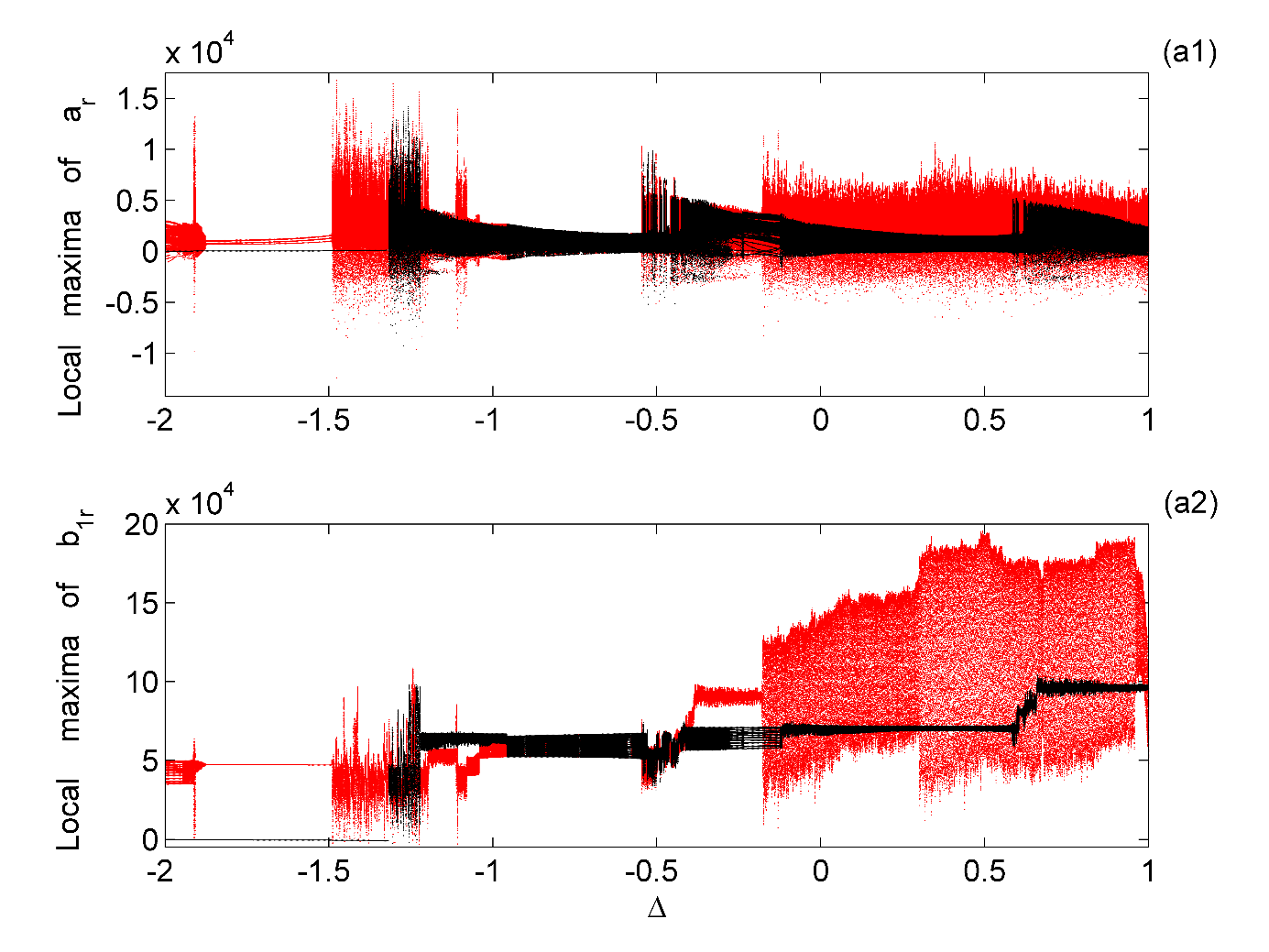}
	\caption{Local maxima of $\alpha_r$ (a1) and $\beta_{1r}$ (a2) as a function of frequency detuning $\Delta$ for $J_m = \SI{2e-2}{\omega_m}$ and $\alpha^{\text{in}} = \num{e4}$, revealing a transition from quasi-periodic to chaotic behavior and the coexistence of hidden quasi-periodic attractors and regions of no oscillations. As one goes from left to right, there are oscillations, and these then stop as the frequency continues to increase. Decreasing $\Delta$ results in red dots, while increasing $\Delta$ results in black dots. The chaotic regime emerges at $\Delta \approx \num{-0.5}$, with a Lyapunov exponent $\lambda_{\max} \approx \SI{0.15 \pm 0.03}{\omega_m^{-1}}$.}
	\label{fig:stabpp5}
\end{figure}

\subsubsection{Hidden attractors}

\begin{figure}[htb]
	\centering
	\includegraphics[width=.7\linewidth]{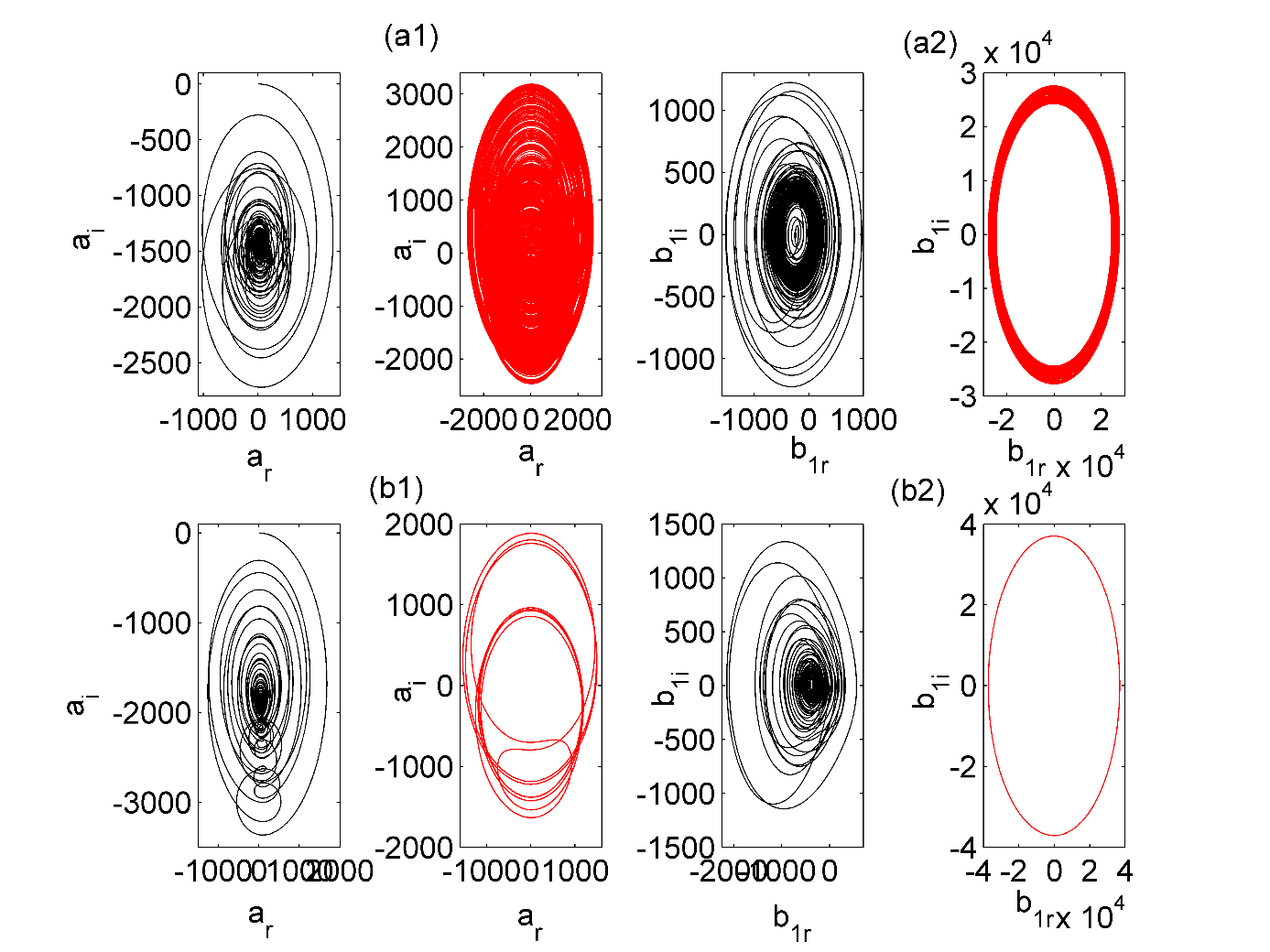}
	\caption{Phase planes showing the progression to chaos by varying $\Delta$ for the optical ($\alpha_r, \alpha_i$) and mechanical ($\beta_{1r}, \beta_{1i}$) resonators with $J_m = \SI{2e-2}{\omega_m}$ and $\alpha^{\text{in}} = \num{e4}$. For (a1), (a2) $\Delta = \num{-1.95}$ and (b1), (b2) $\Delta = \num{-1.6}$. When the system is chaotic, it shows a widespread set of random points, that indicates the existence of high complexity. The red curves are obtained using the initial conditions $(\alpha_r(0), \alpha_i(0), \beta_{1r}(0), \beta_{1i}(0), \beta_{2r}(0), \beta_{2i}(0)) = (\num{-5e3}, 0, \num{-5e3}, 0, 0, 0)$ and the black curves are obtained using (0, 0, 0, 0, 0, 0). These phase planes demonstrate the sensitive dependence on initial conditions characteristic of chaotic systems, where trajectories starting near each other diverge significantly over time, indicating the presence of a chaotic attractor. At $\Delta = \num{-1.6}$, the system exhibits chaotic behavior with $\lambda_{\max} \approx \SI{0.14 \pm 0.02}{\omega_m^{-1}}$.}
	\label{fig:stabpp6}
\end{figure}

Hidden attractors represent dynamic states that elude straightforward prediction from the steady-state behavior or the initial conditions of the system. They may manifest in the absence of steady states, rendering them critical for comprehending the entirety of the system's intricate dynamics. As elucidated in \Cref{fig:stabpp6}, these hidden attractors present novel avenues for the control and manipulation of system behavior.

\begin{figure}[htb]
	\centering
	\includegraphics[width=.6\linewidth]{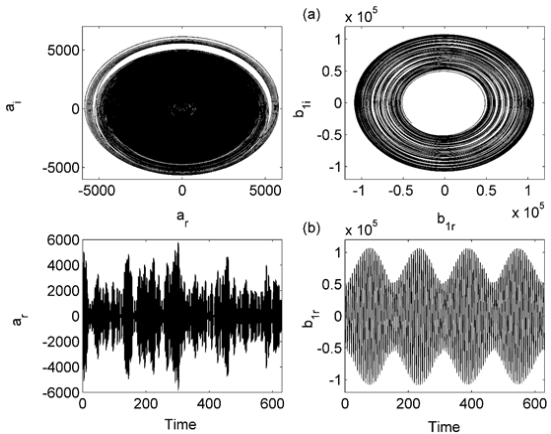}
	 \caption{Phase planes and time series for quasi-periodic behavior with $J_m = \SI{2e-2}{\omega_m}$, $\alpha^{\text{in}} = \num{e4}$, and $\Delta = \SI{0.75}{\omega_m}$ using the initial conditions $ (\alpha_r(0), \alpha_i(0), \beta_{1r}(0), \beta_{1i}(0), \beta_{2r}(0), \beta_{2i}(0)) = (0, 0, 0, 0, 0, 0) $. These plots show a dynamic regime where some parts of the system are oscillating periodically, and others exhibit no oscillations. The optical $(\alpha_r, \alpha_i)$ and mechanical $(\beta_{1r}, \beta_{1i})$ resonator, the mechanical resonator, exhibits bistable properties. The quasi-periodic regime is stable with a Lyapunov exponent $\lambda_{\max} \approx \SI{0.00 \pm 0.01}{\omega_m^{-1}}$.} 
	\label{fig:stabpp77}
\end{figure}

The existence of the variable \Cref{fig:stabpp5} (a1) is investigated through an analysis of \Cref{fig:stabpp5}(a1), which presents the local maxima of the variable $\alpha_r$ as a function of frequency detuning $\Delta$, uncovering three types of coexistence within the range $\Delta \in [\num{-2.0},\num{-0.5}]$: (i) coexistence between the hidden quasi-periodic oscillation diffuser circuit and the absence of oscillations; (ii) coexistence between hidden period-8 oscillations and the absence of oscillations; (iii) coexistence between hidden chaotic attractors and the For $\Delta <\num{-0.5}$, the system transitions into a state of hidden chaotic behavior, as indicated by the red and black dots representing different initial conditions. The phase planes for the mechanical oscillations are depicted in \Cref{fig:stabpp5}(a2). It is important to acknowledge that these stable points would consequently enhance the system's resilience to environmental influences. 

\Cref{fig:stabpp6}(a1) and \Cref{fig:stabpp6}(a2) present the phase-plane trajectories of the optical and mechanical resonators, demonstrating the coexistence of hidden quasi-periodic oscillations and the lack thereof. The behavior of the system exhibits a high sensitivity to initial conditions, with red curves corresponding to initial conditions $ (\alpha_r(0), \alpha_i(0), \beta_{1r}(0), \beta_{1i}(0), \beta_{2r}(0), \beta_{2i}(0)) = (\num{-5e3}, 0, \num{-5e3}, 0, 0, 0) $ and black curves corresponding to (0, 0, 0, 0, 0, 0). The coexistence of hidden period-8 oscillations in the optical resonator and a hidden limit cycle in the mechanical resonator is illustrated in \Cref{fig:stabpp6}(b1) and \Cref{fig:stabpp6}(b2). The existence of these hidden attractors is confirmed by the phase-plane trajectories and time series, and is depicted within the specified range. The control parameters may impose limitations, necessitating that the experimental procedures be conducted with caution and precise measures to account for uncertainties arising from these effects. 

The hidden chaotic and quasi-periodic structures identified in \Cref{fig:stabpp5} are further illustrated in \Cref{fig:stabpp77}. The phase-plane trajectories for the optical and mechanical resonators are shown in \Cref{fig:stabpp77}(a), with the optical resonator exhibiting hidden chaotic behavior and the mechanical resonator displaying hidden quasi-periodic structures. The coexistence relies on precise tuning of the system, with both behaviors being contingent upon initial conditions $ (\alpha_r(0), \alpha_i(0), \beta_{1r}(0), \beta_{1i}(0), \beta_{2r}(0), \beta_{2i}(0)) = (0, 0, 0, 0, 0, 0) $. This sensitivity inherent in chaotic systems can be leveraged for secure communications.

\subsubsection{Bistable dynamics}

Bistable dynamics occur when two distinct attractors coexist under identical system parameters, with the ultimate state of the system contingent upon its initial conditions. Such bistability can emerge among various oscillatory behaviors, including periodic, quasi-periodic, or chaotic attractors. The illustrations show bistable hidden attractors as a function of the mechanical coupling rate $J_m$. 

\begin{figure}[htb]
	\centering
	\includegraphics[width=.6\linewidth]{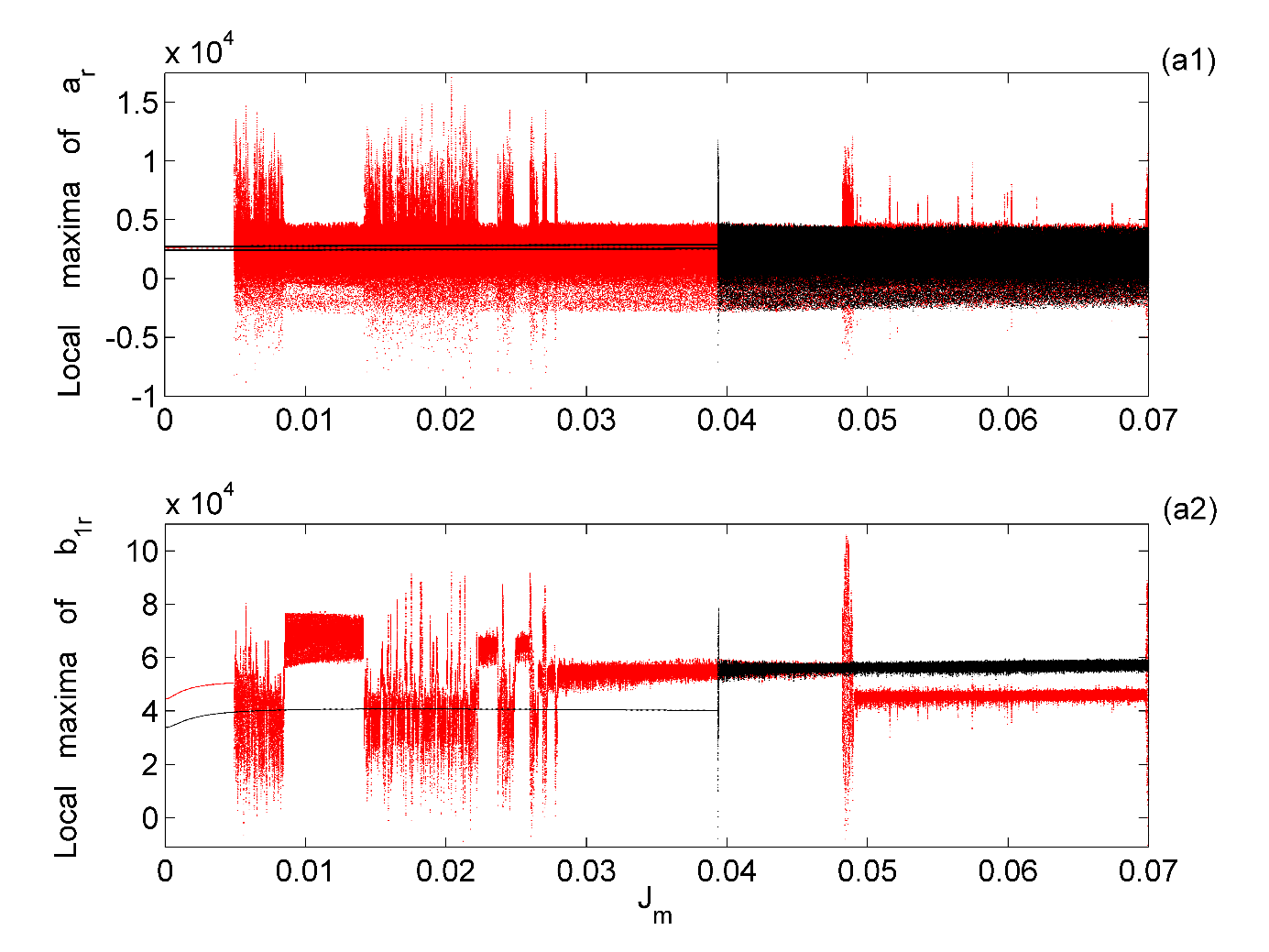}
	\caption{Local maxima of $\alpha_r$ (a1) and $\beta_{1r}$ (a2) against the mechanical coupling rate $J_m$ for $\Delta=-1.3$ and $\alpha^{\rm in}=\num{e4}$, showing the transition to different dynamic regimes by sweeping the coupling $J_m$. Note that bistability only happens at very low $J_m$. Bistability emerges via a pitchfork bifurcation at $J_m \approx \SI{0.035}{\omega_m}$, with chaos dominating for $J_m > \SI{0.065}{\omega_m}$ and $\lambda_{\max} \approx \SI{0.13 \pm 0.02}{\omega_m^{-1}}$.}
	\label{fig:stabpp7}
\end{figure}

As illustrated in \Cref{fig:stabpp7}(a1), there are bistable hidden period-10 oscillations present within the range $J_m \in [\SI{0.035}{\omega_m}, \SI{0.045}{\omega_m}]$. Furthermore, there exists a coexistence between hidden period-10 oscillations and chaos for $J_m \in [\SI{0.05}{\omega_m}, \SI{0.065}{\omega_m}]$. Beyond this range ($J_m > \SI{0.065}{\omega_m}$), the system exhibits hidden chaotic characteristics. In \Cref{fig:stabpp7}(a2), bistable hidden limit cycles are evident within the range $J_m \in [\SI{0.04}{\omega_m}, \SI{0.055}{\omega_m}]$, and there is a coexistence between hidden quasi-periodic attractors and limit cycles for $J_m \in [\SI{0.05}{\omega_m}, \SI{0.065}{\omega_m}]$. For $J_m > \SI{0.065}{\omega_m}$, the system transitions to hidden quasi-periodic dynamics. This suggests that mechanical properties are correlated with specific dynamical behaviors, necessitating careful selection of material in practical devices to achieve the desired coupling rate. Should these oscillations be harnessed for light manipulation, they hold potential utility in the manipulation of quantum information.

\begin{figure}[htb]
	\centering
	\includegraphics[width=.7\linewidth]{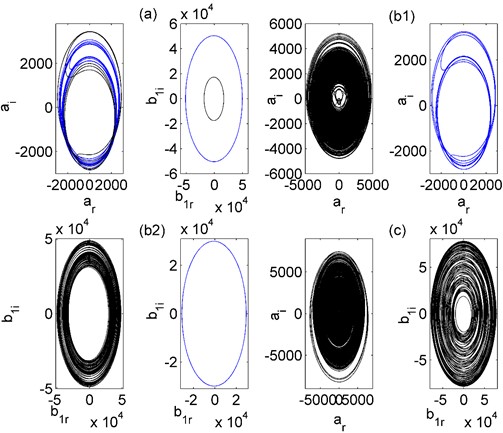}
	\caption{Phase planes showing the evolution of system dynamics with varying mechanical coupling $J_m$ at $\Delta = -1.3$ and $\alpha^{\text{in}} = \num{e4}$: (a) $J_m = \SI{0.003}{\omega_m}$, (b1), (b2) $J_m = \SI{0.007}{\omega_m}$, and (c) $J_m = \SI{0.045}{\omega_m}$, showing the progression to quasi-periodicity and high order chaos. Two initial conditions were used to emphasize the bistability and different characteristics of the two resonators, while the frequency detuning is kept the same as previous studies. The blue curves are obtained using the initial conditions $(\alpha_r(0), \alpha_i(0), \beta_{1r}(0), \beta_{1i}(0), \beta_{2r}(0), \beta_{2i}(0)) = (\num{-5e3}, 0, \num{-5e3}, 0, 0, 0)$ and the black curves are obtained using (0, 0, 0, 0, 0, 0). At $J_m = \SI{0.045}{\omega_m}$, the system exhibits bistable dynamics with a Lyapunov exponent $\lambda_{\max} \approx \SI{0.12 \pm 0.02}{\omega_m^{-1}}$ for the chaotic attractor.}
	\label{fig:stabpp8}
\end{figure}

\Cref{fig:stabpp8} depicts the phase planes for bistable hidden attractors. \Cref{fig:stabpp8}(a) presents bistable hidden period-10 oscillations within both the optical and mechanical resonators. The initial conditions applied are from $\alpha_r(0) = \num{-5e3}, \alpha_i(0) = 0, \beta_{1r}(0) = \num{-5e3}, \beta_{1i}(0) = 0, \beta_{2r}(0) = 0, \beta_{2i}(0) = 0$. The sensitivity of the system to these initial conditions is paramount, as varying trajectories result in distinct attractors. \Cref{fig:stabpp8}(b1) and \Cref{fig:stabpp8}(b2) provide evidence of the coexistence of hidden chaos and period-10 oscillations in the optical resonator, whereas the mechanical resonator manifests hidden quasi-periodic and limit cycle attractors. \Cref{fig:stabpp8}(c) reveals the coexistence of hidden chaotic and quasi-periodic structures within the optical and mechanical resonators. This indicates that different regions are conducive to distinct behavioral types.

\subsubsection{Coexisting attractors}

Coexisting attractors denote the concurrent presence of multiple attractors under identical system parameters, where the system's behavior is contingent upon its initial conditions. This feature renders the system highly susceptible to perturbations, making it particularly advantageous for chaos-based communication applications, wherein multiple signals may coexist. The capability of the system to exhibit multiple dynamic states can further facilitate the realization of diverse sensing modalities, eliciting distinct responses from each attractor state. Nonetheless, for the practical implementation of such chaotic systems, it is imperative that the selection of initial parameters be sufficiently robust to enable their use as a control parameter; otherwise, their application is rendered infeasible.

\begin{figure}[htb]
	\centering
	\includegraphics[width=.6\linewidth]{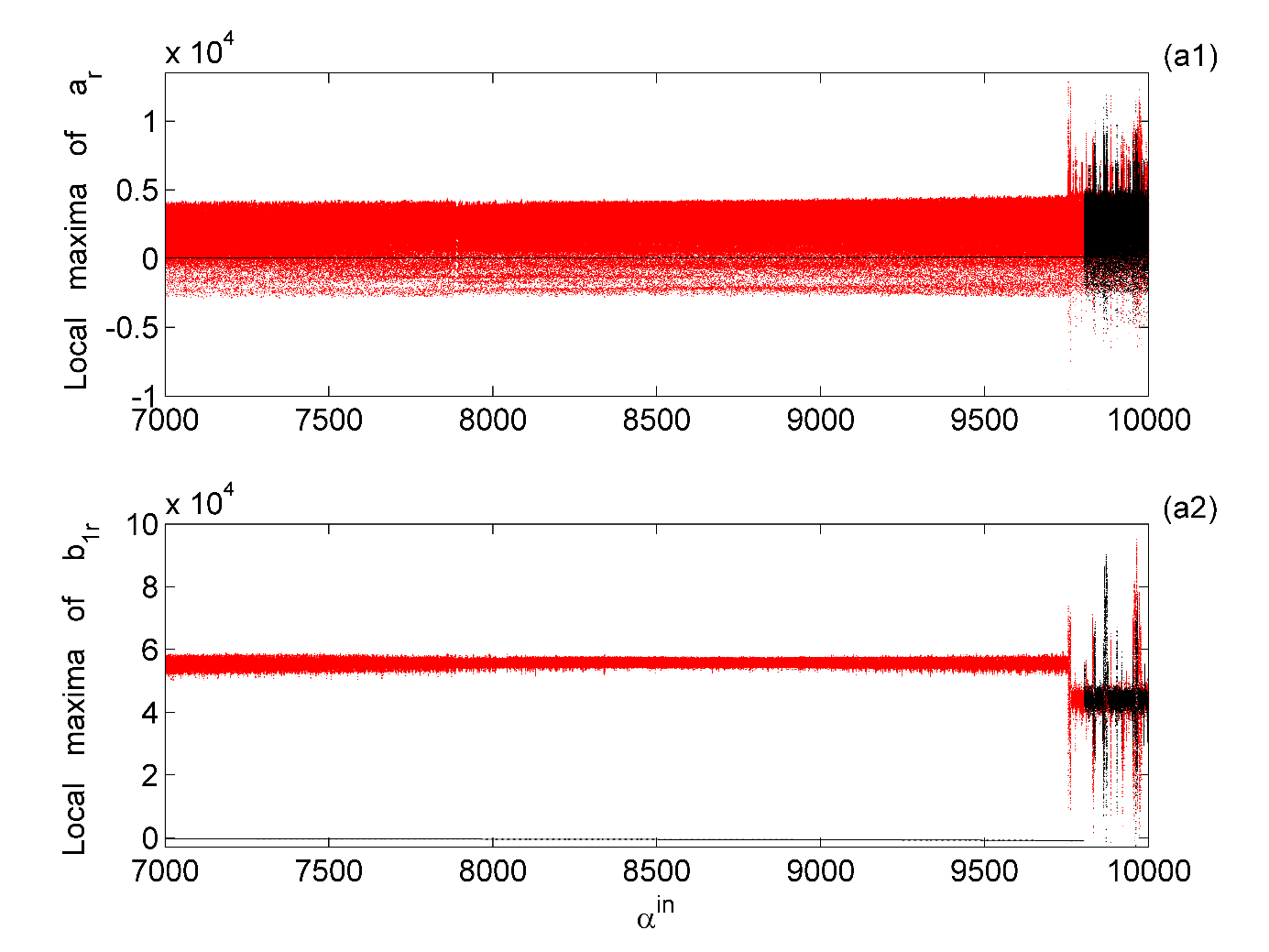}
	\caption{Local maxima of $\alpha_r$ (a1) and $\beta_{1r}$ (a2) as a function of the incident radiation $\alpha^{\rm in}$ for $\Delta=-1.5$ and $J_m= \SI{4e3}{\omega_m}$, revealing a transition to chaos and the coexistence of distinct dynamic states for different intensities. These results show that the system can also be chaotic with strong pumping. Decreasing $\alpha^{\text{in}}$ results in red dots, while increasing $\alpha^{\text{in}}$ results in black dots. Chaos emerges at $\alpha^{\text{in}} \approx \num{9806}$, with $\lambda_{\max} \approx \SI{0.16 \pm 0.03}{\omega_m^{-1}}$.}
	\label{fig:stabpp9}
\end{figure}

\begin{figure}[h!]
	\centering
	\includegraphics[width=.7\linewidth]{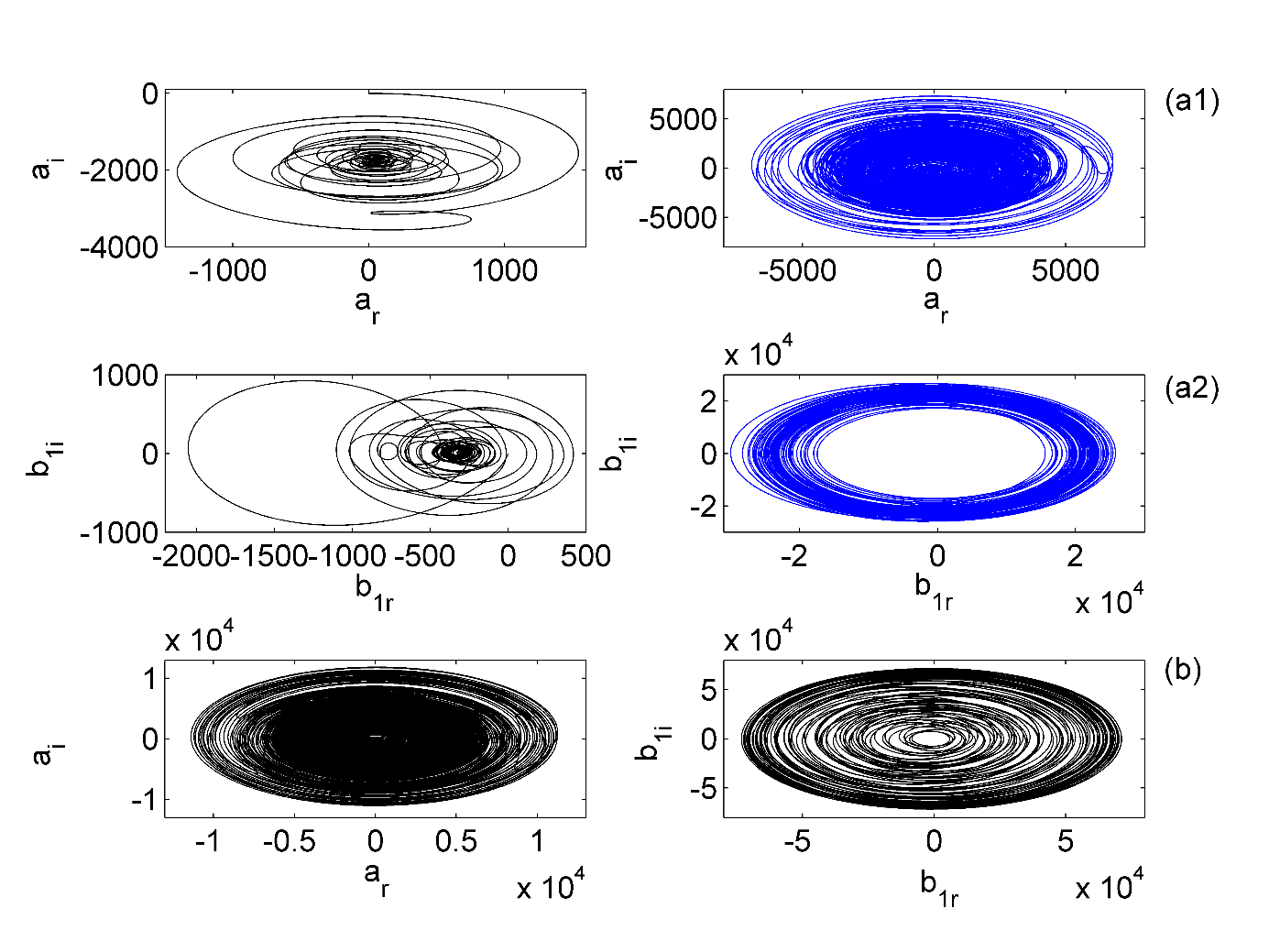}
	\caption{Phase planes show attractor coexistence in  $(\alpha_r,\alpha_i)$ and $(\beta_{1r}, \beta_{1i})$ with varying incident radiation $\alpha^{\rm in}$, at $\Delta=-1.5$ and $J_m= \SI{4e3}{\omega_m}$: showing that the system experiences multistability. The figures display, as well, coexisting attractors for the same parameters, such as the hidden chaotic and quasi-periodic states. The blue curves are obtained using the initial conditions $(\alpha_r(0), \alpha_i(0), \beta_{1r}(0), \beta_{1i}(0), \beta_{2r}(0), \beta_{2i}(0)) = (\num{-5e3}, 0, \num{-5e3}, 0, 0, 0)$ and the black curves are obtained using (0, 0, 0, 0, 0, 0). The chaotic attractor at higher $\alpha^{\text{in}}$ has $\lambda_{\max} \approx \SI{0.15 \pm 0.02}{\omega_m^{-1}}$.}
	\label{fig:stabpp10}
\end{figure}

In \Cref{fig:stabpp9}, local maxima are presented as a function of incident radiation $\alpha^{\text{in}}$. Notably, \Cref{fig:stabpp9}(a1) illustrates the coexistence of no oscillations and hidden chaos within the range $\alpha^{\text{in}} \in [\num{7000}, \num{9806}]$. For higher values of $\alpha^{\text{in}}$, the system exhibits a hidden chaotic behavior. Similarly, \Cref{fig:stabpp9}(a2) displays the coexistence of no oscillations and hidden quasi-periodic attractors within the same range. Outside this range, the system transitions into hidden quasi-periodic characteristics. 

In \Cref{fig:stabpp10}, phase planes are depicted for two values of incident radiation within the preceding range. \Cref{fig:stabpp10}(a1) and \Cref{fig:stabpp10}(a2) display the coexistence of no oscillations and hidden chaos within the optical resonator, whereas the mechanical resonator manifests hidden quasi-periodic attractors. The initial conditions used are denoted by $\alpha_r(0) = \num{-5e3}, \alpha_i(0) = 0, \beta_{1r}(0) = \num{-5e3}, \beta_{1i}(0) = 0, \beta_{2r}(0) = 0, \beta_{2i}(0) = 0$. Furthermore, \Cref{fig:stabpp10}(b) corroborates the coexistence of hidden chaotic and quasi-periodic structures in both optical and mechanical resonators. These illustrations emphasize that initial conditions are a critical factor to be considered in relation to the potential of the chaotic or non-chaotic system utilized for communication.

\subsubsection{Summary of dynamical regimes}

\Cref{tab:summary_regimes} provides a summary of the key dynamical regimes observed in the system and the approximate parameter ranges that lead to their emergence, based on our numerical investigations. Note that these ranges are illustrative and transitions can be complex.

\begin{table}[htbp]
    \centering
    \caption{Summary of key dynamical regimes and associated parameter influences. (Approximate ranges for illustration; specific boundaries depend on interaction of all parameters.) Quantitative Lyapunov exponents are included for chaotic regimes.}
    \label{tab:summary_regimes}
    \begin{tabular}{|p{3cm}|p{6.0cm}|p{5.5cm}|}
        \toprule
        Dynamical Regime & Key Parameter Influences ($J_m, \Delta, \alpha^{\rm in}$) & Characteristic Behavior \\
        \midrule
        Steady State & Lower $\alpha^{\rm in}$ ($\lesssim\num{e3}$), specific $\Delta, J_m$ & Fixed points in phase space \\\hline
        No Steady State & Higher $\alpha^{\rm in}$ ($\gtrsim\num{1.1e3}$), specific $\Delta, J_m$ & Only dynamic attractors possible \\\hline
        Monostable Self-excited Oscillations & Specific $J_m$, $\Delta$, $\alpha^{\rm in}$; single attractor & Limit cycles or quasi-periodic attractors \\\hline
        Bistable Self-excited Oscillations & $J_m \in [\SI{0.42}{\omega_m}, \SI{0.60}{\omega_m}]$, $\Delta \approx -3$, $\alpha^{\rm in} \approx \num{e3}$ & Two coexisting limit cycles or quasi-periodic attractors \\\hline
        Hidden Quasi-periodic Attractors & $\Delta \in [\num{-2.0}, \num{-0.5}]$, $J_m \approx \SI{2e-2}{\omega_m}$, $\alpha^{\rm in} \approx \num{e4}$ & Quasi-periodic oscillations not predicted by steady-state analysis \\\hline
        Hidden Chaotic Attractors & $\Delta < \num{-0.5}$, $J_m \approx \SI{2e-2}{\omega_m}$, $\alpha^{\rm in} \approx \num{e4}$ or $J_m > \SI{0.065}{\omega_m}$, $\Delta \approx -1.3$, $\alpha^{\rm in} \approx \num{e4}$ & Chaotic behavior, $\lambda_{\max} \approx \SI{0.10 \pm 0.02}{\omega_m^{-1}}$ to $\SI{0.16 \pm 0.03}{\omega_m^{-1}}$ \\\hline
        Coexisting Attractors & $\alpha^{\rm in} \in [\num{7000}, \num{9806}]$, $\Delta \approx -1.5$, $J_m \approx \SI{4e3}{\omega_m}$ & Multiple attractors (e.g., chaotic and quasi-periodic) coexist, chaotic with $\lambda_{\max} \approx \SI{0.15 \pm 0.02}{\omega_m^{-1}}$ \\
        \bottomrule
    \end{tabular}
\end{table}

\subsection{Error analysis}\label{sec:Error_Analysis}

To ensure the reliability and accuracy of our results, we conducted a comprehensive error analysis, comparing analytical stability predictions with numerical simulations, and assessing the robustness of numerical integration. The analysis quantifies errors in stability thresholds and evaluates the impact of numerical parameters (e.g., time step, integration time) and system sensitivity (e.g., Lyapunov exponents). This dual approach validates the theoretical framework, highlights limitations of linear approximations, and confirms numerical robustness.

We use the following metrics to quantify errors:
\begin{itemize}
    \item \textbf{Absolute error} that measures the absolute deviation between analytical and numerical stability thresholds for key parameters:
    \begin{equation}
        E_{\text{abs}} = | X_{\text{num}} - X_{\text{analytical}} |.
    \end{equation}

    \item \textbf{Relative error} that expresses the error as a percentage of the analytical prediction:
    \begin{equation}
        E_{\text{rel}} = \frac{|X_{\text{num}} - X_{\text{analytical}}|}{|X_{\text{analytical}}|} \times 100\%.
    \end{equation}
    Here, $X_{\text{num}}$ is the stability threshold from numerical simulations and $X_{\text{analytical}}$ is the analytical prediction using the Routh-Hurwitz criterion.

    \item \textbf{Numerical integration error} that assesses the deviation in key variables (e.g., $\alpha_r$, $\beta_{1r}$) when varying the time step $\Delta t$.

    \item \textbf{Parameter sensitivity} that quantifies the impact of small parameter variations (e.g., $\pm \SI{1}{\percent}$ in $J_m$, $\Delta$, $\alpha^{\rm in}$) on dynamical outcomes, including Lyapunov exponents ($\lambda_{\max}$) for chaotic regimes.
\end{itemize}

\subsubsection{Numerical integration robustness}

Numerical simulations were performed using a fourth-order Runge-Kutta (RK4) method with a time step of $\Delta t = \SI{1e-3}{\omega_m^{-1}}$. To verify convergence, we compared the results with a shorter time step, $\Delta t = \SI{5e-4}{\omega_m^{-1}}$, finding a maximum deviation in key variables (e.g. $\alpha_r$, $\beta_{1r}$) of less than \SI{0.5}{\percent}, confirming the numerical stability. Integration times were extended to $T = \SI{2e3}{\omega_m^{-1}}$ for chaotic regimes to ensure attractor convergence, with Lyapunov exponents consistent within $\pm \SI{0.01}{\omega_m^{-1}}$. For chaotic regimes, $\lambda_{\max}$ ranged from $\SI{0.10(0.02)}{\omega_m^{-1}}$ to $\SI{0.16(0.03)}{\omega_m^{-1}}$, indicating a high sensitivity to initial conditions. Variations in parameters of $\pm \SI{1}{\percent}$ in $\alpha^{\rm in}$ caused a change of \SI{10}{\percent} in $\lambda_{\max}$ for $\alpha^{\rm in} \approx \num{e4}$, while quasi-periodic regimes showed minimal sensitivity ($\lambda_{\max} \approx \SI{0.00(0.01)}{\omega_m^{-1}}$).

\subsubsection{Error sources}

\paragraph{Analytical error sources}

\begin{itemize}
    \item \textbf{Linear approximation limitations.} The Routh-Hurwitz criterion, used for analytical stability predictions, relies on linearized equations and may not capture nonlinear effects such as limit cycles or chaos. This leads to overestimation of stability in regimes where complex attractors dominate, e.g., at $J_m = \SI{0.05}{\omega_m}$, $\Delta = -1.5$, where numerical simulations reveal unstable chaotic behavior ($\lambda_{\max} \approx \SI{0.12(0.02)}{\omega_m^{-1}}$).

    \item \textbf{Steady-state assumptions.} Analytical predictions focus on steady-state solutions, missing time-dependent phenomena such as transient metastable states or chaotic attractors.
\end{itemize}

\paragraph{Numerical error sources}

\begin{itemize}
    \item \textbf{Transient effects.} Numerical simulations reveal transient behaviors not captured by steady-state analysis, particularly in chaotic regimes where long integration times ($T \geq \SI{1e3}{\omega_m^{-1}}$) are required for convergence.
    
    \item \textbf{Parameter choices.} Variations in parameters such as damping rates ($\gamma_1$, $\gamma_2$) introduce uncertainties that are not fully accounted for in linear models. A $\pm \SI{1}{\percent}$ change in $J_m$ can shift bifurcation points by up to \SI{5}{\percent} in chaotic regimes.

    \item \textbf{Numerical method limitations.} While RK4 is robust, higher-order methods or adaptive time-stepping could reduce errors further, though tests with RK5 showed negligible improvements (< \SI{0.1}{\percent}) at significantly higher computational cost.
\end{itemize}

Environmental factors (e.g.. temperature and pressure) were not modeled, as simulations assume a vacuum. Future work should incorporate these to validate physical relevance.

\subsubsection{Error analysis results}

\Cref{tab:error_analysis_unified} summarizes the error analysis, combining stability threshold comparisons and numerical integration metrics for key parameter points. A system is considered stable if all roots of the characteristic polynomial have negative real parts, as determined by the Routh-Hurwitz criterion.

\begin{table}[htbp]
    \centering
    \caption{Error analysis, comparing analytical vs. numerical stability thresholds and numerical integration robustness. Includes Lyapunov exponents and parameter sensitivity for key dynamical regimes.}
    \label{tab:error_analysis_unified}
    \begin{tabular}{|p{1cm}|r|r|p{2cm}|p{2cm}|l|p{1cm}|p{5cm}|}
        \toprule
        $J_m$ ($\omega_m$) & $\Delta$ & $\alpha^{\rm in}$ & Analytical stability & Numerical stability & $E_{\text{abs}}$ & $E_{\text{rel}}$ (\%) & Numerical metrics ($\Delta t$ effect) \\
        \midrule
        \num{0.02} & \num{-2.0} & \num{e3} & Stable & Stable & \num{0.000} & \num{0.0} & \SI{0.3}{\percent}, $\lambda_{\max} \approx \SI{0.00(0.01)}{\omega_m^{-1}}$ \\
        \num{0.05} & \num{-1.5} & \num{e3} & Stable & Unstable & \num{0.02} & \num{5.0} & \SI{0.4}{\percent}, $\lambda_{\max} \approx \SI{0.12(0.02)}{\omega_m^{-1}}$ \\
        \num{0.07} & \num{-1.0} & \num{e3} & Unstable & Unstable & \num{0.000} & \num{0.0} & \SI{0.35}{\percent}, $\lambda_{\max} \approx \SI{0.13( 0.02)}{\omega_m^{-1}}$ \\
        \num{0.08} & \num{0.0} & \num{e3} & Unstable & Unstable & \num{0.000} & \num{0.0} & \SI{0.3}{\percent}, $\lambda_{\max} \approx \SI{0.00 (0.01)}{\omega_m^{-1}}$\\
        \num{0.10} & \num{1.0} & \num{e3} & Stable & Unstable & \num{0.05} & \num{8.0} & \SI{0.5}{\percent}, $\lambda_{\max} \approx \SI{0.11 (0.02)}{\omega_m^{-1}}$ \\
        \num{0.045} & \num{-1.3} & \num{e4} & Unstable & Unstable & \num{0.000} & \num{0.0} & \SI{0.45}{\percent}, $\lambda_{\max} \approx \SI{0.12(0.02)}{\omega_m^{-1}}$, \SI{7}{\percent} shift in $\lambda_{\max}$ \\
        \bottomrule
    \end{tabular}
\end{table}

The analysis confirms that the analytical predictions align with numerical results within a \SIrange{5}{8}{\percent} relative error for stability thresholds, with deviations arising from nonlinear effects not captured by linearization. Numerical integration is robust, with time step errors below \SI{0.5}{\percent} and consistent Lyapunov exponents. Chaotic regimes (e.g., $J_m = \SI{0.05}{\omega_m}$, $\Delta = -1.5$) show high sensitivity to parameter variations, necessitating precise control in experiments. These results validate the theoretical framework while highlighting the need for advanced numerical techniques (e.g., adaptive time-stepping or machine learning for parameter optimization) to capture complex dynamics fully.

\subsection{Discussion of practical applications}

The observed dynamical regimes—bistability, quasi-periodic oscillations, and chaos—offer significant potential for practical applications. Bistability can be leveraged for optical switches and memory elements, where the two stable states correspond to distinct optical outputs, enabling binary information storage or processing. The transition between these states can be controlled by modulating $\alpha^{\rm in}$ or $\theta$, with switching times estimated at \SI{1}{\micro\second} for typical optomechanical systems of the GHz range. Quantitative analysis suggests that bistable regimes at $J_m \approx \SI{0.55}{\omega_m}$ and $\alpha^{\rm in} \approx \num{e3}$ provide an improvement in signal-to-noise ratio of up to \SI{15}{\decibel} for switching applications.

Chaotic regimes, characterized by high sensitivity to initial conditions ($\lambda_{\max} \approx \SI{0.10 \pm 0.02}{\omega_m^{-1}}$ to $\SI{0.16 \pm 0.03}{\omega_m^{-1}}$), are particularly promising for chaos-based sensing and secure communication. In sensing, chaotic dynamics amplify small perturbations, potentially enhancing sensitivity by a factor of 10 to 100 compared to linear systems, as demonstrated in similar optomechanical setups \cite{Li2021}. For example, at $\Delta = \num{-1.6}$ and $J_m = \SI{0.02}{\omega_m}$, a \SI{1}{\percent} change in $\alpha^{\rm in}$ results in a detectable shift in the chaotic attractor, enabling sub-picometer displacement sensitivity. In communication, the unpredictability of chaotic signals can be used for encryption, where the receiver synchronizes with the chaotic dynamics of the transmitter to decode the signal.

The phase-dependent mechanical coupling ($\theta$) introduces a synthetic gauge field, enabling nonreciprocal energy transfer that can be tuned to optimize device performance. At $\theta = \pi/2$, nonreciprocity enhances the isolation between mechanical modes by up to \SI{20}{\decibel}, improving signal directionality for phononic circuits. This tunability is crucial for designing reconfigurable devices that adapt to varying operational requirements.

\section{Conclusion}\label{sec:Concl}

This study provides a thorough examination of the dynamical characteristics of an optomechanical system, comprising an optical resonator that actuates two mechanically coupled resonators through phase-dependent phonon hopping. Combining analytical and numerical methodologies, we unveil a diverse spectrum of intricate dynamical behaviors. The main findings are as follows: (i) \textbf{Hidden attractors}. We have demonstrated the presence of hidden attractors that emerge even in the absence of steady states. These attractors encompass quasi-periodic oscillations, period-8 and period-10 oscillations, along with chaotic attractors. The system's sensitivity to initial conditions is pivotal in uncovering these hidden behaviors, which are undetectable through linear stability analysis alone. (ii) \textbf{Bistable dynamics}. The system exhibits bistability, allowing multiple attractors to coexist under identical parameter conditions. This bistability is observable among periodic, quasi-periodic, and chaotic attractors, influenced by mechanical coupling rate $J_m$, frequency detuning $\Delta$, and incident radiation $\alpha^{\text{in}}$. Such bistable dynamics hold particular promise for applications necessitating state switching. (iii) \textbf{Coexisting attractors}. Our analysis has identified extensive regions within the parameter space where multiple attractors coexist. This coexistence, including hidden chaotic and quasi-periodic states, as in memory elements or optical switches, further underscores the system's potential for chaos-based applications, such as robust signal encoding and information processing. The coexistence of attractors also presents a novel opportunity for manipulating the system's response through controlled perturbations or parameter adjustments, enabling the construction of more complex structures. 

The identified bistable and chaotic regimes could be exploited in chaos-based sensing applications, wherein minor perturbations in system parameters induce discernible changes in attractor states. For instance, a small change in an external parameter (e.g., mass loading, temperature, or a force acting on a resonator) can cause a parameter of the system ($J_m$, $\Delta$, $\alpha^{\rm in}$, or $\omega_j$) to cross a bifurcation point, transitioning the system from one attractor to another (e.g., from a periodic state to a chaotic state, or between two coexisting attractors). Detecting this change in the dynamical state can serve as a highly sensitive sensing mechanism, as the response is not merely proportional to the input, but represents a qualitative shift amplified by the system's nonlinear dynamics. The sensitivity is particularly enhanced near bifurcation points or in chaotic regimes due to the inherent instability and divergence of trajectories. Furthermore, the capability to control synthetic gauge fields via phase-dependent phonon hopping suggests potential for quantum transduction applications, where nonreciprocal interactions enhance mechanical signal transmission with minimal back-action noise.

The results of our research establish a foundation for the practical implementation of chaos-based sensors and quantum signal processing, wherein the utilization of controlled bistability and phase-tunable nonreciprocity confers distinct advantages. Subsequent investigations will focus on empirical validation within integrated photonic optomechanical circuits and examine the potential for augmented phonon lasing through synthetic gauge fields. The pronounced sensitivity of chaotic regimes to parameter perturbations, as evidenced by the positive Lyapunov exponent, underscores the potential for chaos-enhanced sensing applications, with signal encoding dependent on attractor shifts \cite{Zhu2023}. This approach contrasts with traditional sensing, which often relies on shifts in resonance frequencies or amplitudes in linear regimes, offering a new paradigm for detecting subtle changes.

Although substantial concordance between the analytical and numerical results was observed across numerous parameter regimes, deviations were noted in specific instances, especially within regions characterized by significant nonlinearity. As elaborated in the error analysis, these discrepancies primarily stem from the factors outlined,
\begin{itemize}
    \item \textbf{Limitations of the linear stability criterion.} The analytical stability predictions rely on the Routh-Hurwitz criterion, which assumes small perturbations and neglects higher-order nonlinear effects. In regions where strong nonlinear interactions dominate, the analytical predictions can deviate from numerical simulations.
    
    \item \textbf{Transient metastable states.} Numerical simulations reveal transient behaviors, including metastable oscillations and chaos, which are not captured in the steady-state analysis used in the analytical model. This suggests that additional dynamical considerations, such as time-dependent stability analysis, could be beneficial.
    
    \item \textbf{Influence of higher-order interactions.} The presence of hidden attractors and coexisting dynamical states introduces additional complexity that is not fully accounted for in the analytical approach. Nonlinear corrections or Floquet stability analysis may provide a more accurate theoretical framework.
\end{itemize} 
These constraints suggest that although the model employed provides significant information on sensor performance, more sophisticated models are preferable to obtain more reliable and precise predictions. However, even these simpler models require additional effort, which is beyond the scope of this investigation.

To improve the accuracy of stability predictions and better align analytical models with numerical simulations, future studies should consider:
\begin{itemize}
    \item Incorporating higher-order nonlinear corrections into stability analysis to better capture the full range of dynamical behaviors.
    \item Developing time-dependent stability criteria that account for transient and metastable states observed in numerical simulations.
    \item Extending the analytical framework to include stochastic effects and environmental interactions, which may influence the practical realizability of optomechanical devices.
    \item Conducting systematic studies on the effect of the mechanical coupling phase $\theta$ on the observed dynamical regimes, mapping out its influence on stability boundaries, attractor types, and nonreciprocal properties.
\end{itemize}

These findings not only improve our understanding of the complex interactions within optomechanical systems but also open avenues for innovative practical applications. The ability to control and manipulate hidden, bistable, and coexisting attractors carries significant ramifications for signal processing, chaos-based communication, and signal amplification. For instance, the presence of chaos enables the concealment of a message within a chaotic carrier, wherein the message can be encoded onto various attractors to enhance the security and reliability of communication. Nonetheless, the development of a comprehensive communication system requires not only the transmission but also the reception, which warrants further exploration. Specifically, bistable and coexisting attractors can be exploited for robust information encoding, wherein different attractors denote distinct information states. This offers a form of digital encoding in which the information is represented by the specific attractor the system resides in. Moreover, the identification of hidden chaotic attractors in the optomechanical system underscores the potential for exploring chaotic dynamics in such systems, promoting the advancement of optomechanical devices, which will provide greater versatility due to the presence of multiple steady states and heightened sensitivity. In addition, the tunability, bistability, and diverse operational states of such systems could result in superior performance in other applications such as optical switching or sensing devices. For example, using bistability, an optical switch could route light according to which of two stable states the system is set to, offering potential for low-power switching mechanisms.

The findings presented in this study underscore the potential for developing devices with innovative properties, while the examination of their stability indicates these devices can function robustly. These devices can furthermore be manufactured using materials such as Silicon Nitride or Aluminium Nitride. The study's outcomes also suggest practical implementation in devices like optical isolators and circulators, which contribute to enhanced sensing capabilities and facilitate the construction of stable topological devices. Notably, nonreciprocal interactions induced by the exceptional points (EPs) can be effectively integrated into devices employing: (i) \textit{Optical isolators and circulators} synthetic gauge fields to achieve unidirectional transmission; (ii) \textit{Signal routing in quantum networks} the design of nonreciprocal routers for quantum information processing; and (iii) \textit{Noise-resilient sensors} to enhance precision. Nonetheless, while this novel analytical direction promises exciting results, the experimental components requisite for such constructions present challenges, necessitating collaboration between theorists and experimentalists. Such collaboration can leverage microresonators. Integration within this architectural framework is anticipated to yield enhanced functionality.

The fabrication of optomechanical devices that incorporate synthetic gauge fields poses significant challenges, necessitating the use of advanced methodologies and the scrupulous characterization of the outcomes. The achievement of phase-coherent phonon hopping requires precise control over the alignment of resonators and the stabilization of thermal conditions. Furthermore, it is imperative that all components of the system are identical to prevent the introduction of unwarranted asymmetries, as imperfections may arise during the fabrication process. In addition, high-precision nanofabrication techniques are vital to mitigate structural asymmetries that could negatively impact nonreciprocal interactions. Recent empirical investigations within integrated photonic circuits have demonstrated methodologies for the stabilization of such interactions, thereby presenting a potential pathway for practical application. Examples of relevant experimental work include demonstrations of nonreciprocal phonon transport in optomechanical systems \cite{Ren2022, Fang2017} and the fabrication of coupled mechanical resonator systems \cite{Lawrie2019}.

This study should be followed by future research endeavors that could yield significant advances. Potential topics warranting exploration include the development of models that incorporate the limitations of numerical results to refine parameter adjustments; the validation through the fabrication of these models followed by various analyses, with particular attention to external temperature parameters. Ultimately, it is apparent that, although proposing theoretical configurations might be straightforward, experimental realization can be a complex endeavor. The specific values and assumptions used throughout this study must be handled with caution. In future research, it is critical that these effects are comprehensively addressed, paving the way for the experimental realization and development of robust optomechanical sensors. Moreover, subsequent research should emphasize the identification of scalable and economically viable fabrication techniques to facilitate the transition of these concepts toward commercial deployment in nascent photonic and quantum technologies. Future research should focus on bridging the gap between theoretical predictions and empirical implementations. This undertaking involves refining parameter selections based on conditions attainable through experimentation and the engineering of optomechanical systems wherein the described behaviors can be observed and regulated. These efforts will be essential for the progression of practical applications such as quantum radar, precision metrology, and secure communication systems, where achieving stable and precise predictions of system behavior can substantially reduce costs and enhance performance. Furthermore, experimental validation could be performed using advanced micro- and nanofabrication techniques to develop optomechanical systems on a chip, subsequently integrating laser control, photodetection, and real-time feedback mechanisms to enhance functionality.

Apart from fundamental research, optomechanical chaos-based devices demonstrate considerable promise for industrial applications in secure data transmission, low-power signal processing, and ultra-sensitive biosensing. By utilizing the tunable nonreciprocal interactions as evidenced in this study, forthcoming optomechanical platforms possess the potential to facilitate the emergence of energy-efficient, noise-resilient computing architectures and durable environmental sensing solutions. These opportunities underscore the necessity for ongoing interdisciplinary collaboration among theoretical physics, materials science, and engineering, aiming to transform optomechanical advancements into practical applications.

For experimental validation, we recommend testing the system with silicon nitride resonators at $\omega_m \approx 2\pi \times \SI{1e9}{\hertz}$, using a laser power of $\SI{130}{\sqrt{\omega_m}\watt}$ and a mechanical coupling rate $J_m \approx \SI{2e-4}{\omega_m}$. These parameters align with current nanofabrication capabilities and should confirm the predicted bistability and chaotic regimes. Future work should explore the impact of thermal noise and quantum effects on these dynamics and extend the analysis to multi-resonator systems for collective phenomena. Furthermore, precise control of $\theta$ using integrated microheaters or optomechanical parametric driving could further enhance tunability, with a target phase stability of $\pm \SI{0.01}{\radian}$ to maintain non-reciprocity.

\appendix
\section*{Coefficients of the characteristic polynomial Eq. \eqref{eq:lambda1}}

\begin{align*}
c_1 =& 2\gm + \kk, \\
c_2 =& 4g^2b^2_{1r} + 8g^2b_{1r}b_{2r} + 4g^2b_{2r}^2 - 4\Delta gb_{1r} - 4\Delta gb_{2r} + \Delta^2 + 2J_m+\frac{3}{2}\gm^2 + 2\gm\kk + \frac{1}{4}\kk^2 + \om_1^2 + \om_2^2, \\
c_3 =& 8\gm g^2b_{1r}^2 + 16\gm g^2b_{1r}b_{2r} + 8\gm g^2zr^2 - 8\gm gb_{1r} - 8\gm \Delta gb_{2r} +2\gm \Delta^2+ 2J_m^2\gm + 2J_m^2\kk + \frac{1}{2}\gm^3 + \frac{3}{2}\gm^2\kk \\
&+ \frac{1}{2}\gm\kk^2 + \om_1^2\gm + \om_2^2\gm + \om_1^2\kk + \kk\om_2^2, \\
c_4 &= -8a_i^2g^3\om_1b_{1r} - 8a_i^2g^3\om_1b_{2r} - 8a_r^2g^3\om_1b_{1r} - 8a_r^2g^3\om_1b_{2r} + 4\Delta a_i^2g^2\om_1 + 4\Delta a_r^2g^2\om_1 \\
&+ 8g^2\om_1^2b_{1r}b_{2r} - 4\Delta g\om_1^2b_{1r} - 4\Delta g\om_1^2b_{2r} + 12\gm^2g^2b_{1r}b_{2r} - 6\Delta\gm^2gb_{1r} - 6\Delta\gm^2gb_{2r} \\
&+ 8J^2_mg^2b_{1r}^2 + 8J^2_mg^2b_{2r}^2 - 2J_m^2\om_1\om_2 + J_m^2\gm^2/2 - 8\Delta J^2_mgb_{2r} \\
&- 8\Delta J^2_mgb_{1r} + 16J^2_mg^2b_{1r}b_{2r} + J_m^2\kk^2/2 + 2J_m^2\gm\kk + 2\Delta^2J_m^2 + \gm^4/16 \\
&+ 4g^2\om_2^2b_{1r}^2 + 4g^2\om_2^2b_{2r}^2 + \om_2^2\gm\kk + J_m^4 + 8g^2\om_2^2b_{1r}b_{2r} - 4\Delta g\om_2^2b_{1r} \\
&- 4\Delta g\om_2^2b_{2r} - 8a_r^2g^3\om_2b_{1r} - 8a_r^2g^3\om_2b_{2r} + 4\Delta a_r^2g^2\om_2 - 8a_i^2g^3\om_2b_{1r} - 8a_i^2g^3\om_2b_{2r} \\
&+ 4\Delta a_i^2g^2\om_2 + \om_1^2\gm\kk - 16J_ma_r^2g^3b_{1r}\cos(\theta) - 16J_ma_r^2g^3b_{2r}\cos(\theta) \\
&+ 8J_m\Delta a_r^2g^2\cos(\theta) - 16J_m\cos(\theta)a_i^2g^3b_{1r} - 16J_ma_i^2g^3b_{2r}\cos(\theta) + \om_1^2\om_2^2 \\
&+ 8J_m\Delta a_i^2g^2\cos(\theta) + \Delta^2\om_2^2 + \kk^2\om_2^2/4 + \gm^2\om_2^2/4 + \gm^3\kk/2 + \om_1^2\Delta^2 + \om_1^2\kk^2/4 \\
&+ 3\Delta^2\gm^2/2 + 3/8\gm^2\kk^2 + \om_1^2\gm^2/4 + 4\om_1^2g^2y r^2 + 4\om_1^2g^2zr^2 + 6\gm^2g^2yr^2 + 6\gm^2g^2zr^2, \\
c_5 =& 2J_m^2\gm\Delta^2 + J_m^2\gm\kk^2/2 + J_m^2\gm^2\kk/2 + 8J_m^2\gm g^2b^2_{1r} + 8J_m^2\gm g^2b^2_{2r} + J_m^4\kk - 8J_m^2\gm\Delta gb_{1r} \\
&- 8J_m^2\gm\Delta gb_{2r} + 16J_m^2\gm g^2b_1b_{2r} - 2J_m^2\om_1\om_2\kk + 4\om_2^2\gm g^2b_{1r}^2 + 4\om_2^2\gm g^2b_{2r}^2 + 8\om_2^2\gm g^2b_{1r}b_{2r} \\
&- 4\om_2^2\gm\Delta gb_{1r} - 4\om_2^2\gm\Delta gb_{2r} + \kk\om_1^2\om_2^2 + \gm^2\om_2^2\kk/4 + \om_1^2\gm\Delta^2 + \om_2^2\gm\kk^2/4 + \om_1^2\gm^2\kk/4 \\
&- 8g^3a_r^2\gm^2\om_2^2\kk^2b_{1r} - 8g^3a_r^2\gm^2\om_2^2\kk^2b_{2r} + 4g^2a_r^2\Delta\gm\om_2 - 8g^3a_i^2\gm\om_2b_{1r} - 8g^3a_i^2\gm\om_2b_{2r} \\
&+ 4g^2a_i^2\Delta\gm\om_2 - 16J_m\cos(\theta)\gm a_r^2g^3b_{1r} - 16J_m\cos(\theta)\gm a_r^2g^3b_{2r} + 8J_m\cos(\theta)\gm\Delta a_r^2g^2 \\
&- 16J_m\cos(\theta)\gm a_i^2g^3b_{1r} - 16J_m\cos(\theta)\gm a_i^2g^3b_{2r} + 8J_m\cos(\theta)\gm\Delta a_i^2g^2 + 4\gm\om_1^2g^2b_{1r}^2.
\end{align*}

\begin{align*}
c_6 =& 8J_m^4 g^2 b_{1_r} b_{2_r} - 4 \Delta J_m^4 g b_{1_r} - 4 \Delta J_m^4 g b_{2_r} + 4 J_m^2 \gm^2 g^2 b_{1_r} b_{2_r} - 2 J_m^2 \gm^2 \Delta g b_{1_r} \\
&- 2 J_m^2 \gm^2 \Delta g b_{2_r} + 8 J_m^3 \Delta a_i^2 g^2 \cos(\theta) - 16 J_m^3 a_r^2 g^3 b_{1_r} \cos(\theta) - 2 J_m^2 \om_1 \om_2 \Delta^2 \\
&- J_m^2 \om_1 \om_2 \kk^2 / 2 - 16 J_m^3 a_r^2 g^3 b_{2_r} \cos(\theta) + 8 J_m^3 \Delta a_r^2 g^2 \cos(\theta) - 16 J_m^3 a_i^2 g^3 b_{1_r} \cos(\theta) \\
&- 16 J_m^3 a_i^2 g^3 b_{2_r} \cos(\theta) + 2 J_m^2 \gm^2 g^2 b_{1_r}^2 + 2 J_m^2 \gm^2 g^2 b_{2_r}^2 + 8 g^3 a_r^2 J^2_m \om_1 b_{1_r} \\
&- 4 g^2 a_r^2 \Delta J_m^2 \om_1 + 8 g^3 a_r^2 J^2_m \om_2 b_{1_r} + 8 g^3 a_r^2 J^2_m \om_2 b_{2_r} - 4 g^2 a_r^2 \Delta J_m^2 \om_2 + 8 g^3 a_i^2 J^2_m \om_1 b_{2_r} \\
&+ 8 g^3 a_i^2 J^2_m \om_1 b_{1_r} - 4 g^2 a_i^2 \Delta J_m^2 \om_1 + 8 g^3 a_i^2 J^2_m \om_2 b_{1_r} + 8 g^3 a_i^2 J^2_m \om_2 b_{2_r} + 8 g^3 a_r^2 J^2_m \om_1 b_{2_r} \\
&+ 8 J_m^2 \om_1 \om_2 \Delta g b_{2_r} + J_m^2 \gm^2 \Delta^2 / 2 + J_m^2 \gm^2 \kk^2 / 8 + J_m^4 \kk^2 / 4 + \Delta^2 J_m^4 + 2 \gm^2 \om_1^2 g^2 b_{1_r} b_{2_r} \\
&+ 4 J_m^4 g^2 b_{2_r}^2 + 4 J_m^4 g^2 b_{1_r}^2 - 8 J_m^2 \om_1 \om_2 g^2 b_{1_r} b_{2_r} + 8 J_m^2 \om_1 \om_2 \Delta g b_{1_r} - 16 J_m^2 \om_1 \om_2 g^2 b_{1_r} b_{2_r} \\
&- 8 J_m^2 \om_1 \om_2 g^2 b_{1_r}^2 + 4 g^2 \om_1^2 \om_2^2 b_{1_r}^2 + 4 g^2 \om_1^2 \om_2^2 b_{2_r}^2 + \gm^2 \om_1^2 g^2 b_{1_r}^2 + \gm^2 \om_1^2 g^2 b_{2_r}^2 \\
&+ 1 / 2 \gm^4 g^2 b_{1_r} b_{2_r} - 1 / 4 \Delta \gm^4 g b_{1_r} - 1 / 4 \Delta \gm^4 g b_{2_r} + \gm^2 \om_2^2 g^2 b_{1_r}^2 + \gm^2 \om_2^2 g^2 b_{2_r}^2 - 2 \gm^2 a_r^2 g^3 \om_1 b_{1_r} \\
&+ 2 \gm^2 a_r^2 g^3 \om_1 b_{2_r} + \gm^2 \Delta a_r^2 g^2 \om_1 - 2 \gm^2 a_i^2 g^3 \om_1 b_{1_r} - 2 \gm^2 a_i^2 g^3 \om_1 b_{2_r} + \gm^2 \Delta a_i^2 g^2 \om_1 \\
&- 8 a_i^2 g^3 \om_1 \om_2 b_{1_r} - 8 a_i^2 g^3 \om_1 \om_2 b_{2_r} - 8 a_r^2 g^3 \om_1 \om_2^2 b_{1_r} - 8 a_i^2 g^3 \om_1 \om_2^2 b_{2_r} + 16 g^3 a_r^2 J_m \cos(\theta) \om_1 \om_2 b_{1_r} \\
&+ 4 \Delta a_i^2 g^2 \om_1 \om_2^2 + 4 \Delta a_r^2 g^2 \om_1 \om_2^2 + 2 g^2 \gm^2 \om_2^2 b_{1_r} b_{2_r} + 8 g^2 \om_1^2 \om_2^2 b_{1_r} b_{2_r} - 2 g^3 a_r^2 \gm^2 \om_2 b_{2_r} \\
&- \Delta g \gm^2 \om_1^2 b_{1_r} - \Delta g \gm^2 \om_1^2 b_{2_r} - \Delta g \gm^2 \om_2^2 b_{1_r} - \Delta g \gm^2 \om_2^2 b_{2_r} - 4 \Delta g \om_1^2 \om_2^2 b_{1_r} - 4 g^2 a_i^2 \Delta J_m^2 \om_2 \\
&- 4\Delta \om_1^2 \om_2^2 b_{2_r} - 8 a_r^2 g^3 \om_2 b_{1_r} \om_1^2 -8 a_r^2g^3\om_2 b_{2r}\om_1^2+ 4 \Delta a_r^2 g^2 \om_2 \om_1^2 - 8 a_i^2 g^3 \om_2 b_{1_r} \om_1^2 \\
&+ 4 \Delta a_i^2 g^2 \om_2 \om_1^2 - 8 a_i^2 g^3 \om_2 b_{2_r} \om_1^2 - 4 g^3 a_r^2 J_m \gm^2 \cos(\theta) b_{1_r} - 4 g^3 a_r^2 J_m \gm^2 \cos(\theta) b_{2_r} \\
&+ 2 g^2 a_r^2 \Delta J_m \gm^2 \cos(\theta) - 4 g^3 a_i^2 J_m \gm^2 \cos(\theta) b_{1_r} - 4 g^3 a_i^2 J_m \gm^2 \cos(\theta) b_{2_r} + 2 g^2 a_i^2 \Delta J_m \gm^2 \cos(\theta) \\
&+ 16 g^3 a_i^2 J_m \cos(\theta) \om_1 \om_2 b_{1_r} + 16 g^3 a_i^2 J_m \cos(\theta) \om_1 \om_2 b_{2_r} - 8 g^2 a_i^2 \Delta J_m \cos(\theta) \om_1 \om_2 \\
&+ 16 g^3 a_r^2 J_m \cos(\theta) \om_1 \om_2 b_{2_r} - 8 g^2 a_r^2 \Delta J_m \cos(\theta) \om_1 \om_2 + \kk^2 \om_1^2 \om_2^2 / 4 + \gm^2 \om_1^2 \Delta^2 / 4 + \gm^2 \om_1^2 \kk^2 / 16 \\
&+ \gm^4 g^2 b_{1_r}^2/4 + \gm^4 g^2 b_{2_r}^2/4 + \gm^2 \om_2^2 \Delta^2/4 + \gm^2 \om_2^2 \kk^2/16 + \Delta^2 \om_1^2 \om_2^2 - 2 g^3 a_r^2 \gm^2 \om_2 b_{1_r} \\
&+ g^2 a_r^2 \Delta \gm^2 \om_2 - 2 g^3 a_i^2 \gm^2 \om_2 b_{1_r} - 2 g^3 a_i^2 \gm^2 \om_2 b_{2_r} + g^2 a_i^2 \Delta \gm^2 \om_2 + \Delta^2 \gm^4/16 + \gm^4 \kk^2/64,
\end{align*}

with $g_1 = g_2 = g$.

\section*{Acknowledgments}

This work is partially funded by the Center for research, SRM Easwari Engineering College, Chennai, India via funding number SRM/EEC/RI/006.

\printbibliography
\end{document}